\documentclass[aps,pre,twocolumn,superscriptaddress,showpacs,showkeys ]{revtex4-1}

\usepackage{graphicx}                   
\usepackage{hyperref}                   
\usepackage{amsmath,amssymb}            
\usepackage{epstopdf}
\usepackage{subcaption}					

\usepackage{color}
\definecolor{orange}{rgb}{1,0.5,0}
\definecolor{brown}{rgb}{0.65, 0.16, 0.16}
\definecolor{phlox}{rgb}{0.87, 0.0, 1.0}

\begin{document}

\title{Fillips in Sandpiles\\
	Spanning Avalanches, Bifurcation and Temporal oscillations}

\author{M. N. Najafi}
\affiliation{Department of Physics, University of Mohaghegh Ardabili, P.O. Box 179, Ardabil, Iran}
\email{morteza.nattagh@gmail.com}

\author{Z. Moghadam}
\affiliation{Department of Physics, University of Mohaghegh Ardabili, P.O. Box 179, Ardabil, Iran}
\email{zahramoghadam.physics@gmail.com }

\begin{abstract}
The manipulation of the self-organized critical systems by repeatedly deliberate local relaxations (fillips) affect considerably the dynamics of avalanches and change their evolution. During a fillip, the energy diffuses to the neighboring regions, causing a smoothening of the height filed over the system. The fillips are controlled by a parameter $\zeta$ which is related to the number of local smoothening events in an avalanche. The system shows a new (mass and time) scales, leading to some oscillatory behaviors. A bifurcation occurs at some $\zeta$ value, above which some oscillations occur in the mean energy, and also in the autocorrelation functions. These oscillations are associated with \textit{spanning avalanches} which are due to the accumulation of energy in the smoothed system. The analysis of the rare event waiting time (REWT) confirms also the appearance of this new time scale. A mean field analysis is presented to explain some results.
\end{abstract}


\maketitle

\section{Introduction}
The critical-state dynamics of avalanches had been the center of attention, since the advent of the self-organized critical (SOC) systems~\cite{BTW,Dhar,Paczuski}. The sandpiles, as a prototype of SOC systems, are slowly driven and evolve (without tuning of any external parameter) towards a steady critical state which is characterized by long-range spatial correlations and power-law behaviors. The non-linearity of this model arises from a threshold value, in such a way that when the local energy in some point of the system exceeds the threshold, the energy starts to spread throughout the sample. The dynamics of the resulting avalanches is governed by the local updates (relaxations) depending on the state of sites at the moment, i.e. a domino (chain of relaxations) runs over the system which is mainly affected by the \textit{minimally stable sites} (MSS)~\cite{BTW}. The latter (MSSs) are defined as the sites which become unstable under a single stimulation, i.e. adding an energy unit. The network comprised by MSSs is determinative in the dynamics of the underlying avalanche. When this network is very dense, i.e. the number of MSSs is high, the signal (caused by local stimulation) propagates throughout the sample and spans nearly whole of the system. Under the overall evolution of the system, other sort of stable sites (i.e. more-than-minimally stable sites) appear in an increasing rate, which impede the propagation of the signal. In the steady state, the mentioned network becomes dilute such that most signals are not spanning, i.e. signals become finite-ranged, and the MSS network becomes self-similar~\cite{BTW}. In this case, the added energies are accumulated in the system, causing some large-scale (rare) events during which a large amount of energy leave the system. The scale-invariant structure of the MSS network induces some critical behaviors which are reflected in power-law behaviors~\cite{Dhar}, $1/f$-noise~\cite{najafiNoise}, conformal invariance~\cite{najafiSLE}, etc. The connection of the system in the critical state to the other statistical models is well-understood and well-studied. The examples are the connection with spanning trees~\cite{Majumdar}, ghost models~\cite{Majumdar}, q-state Potts model~\cite{Saleur}, etc. For a good review, refer to~\cite{Dhar}. \\
The relation of the sandpile models to some natural phenomenon, like earthquake~\cite{Sornette}, fluid propagation in reservoirs~\cite{NajafiWater} and neuronal activities~\cite{Arcangelis}, make the characterization of their critical states worthy to study. One output of such a study may be an understanding and controlling the rare (large-scale) events. Importantly one may ask what the response of avalanches to the external manipulation of the local energy content of the system is. Especially what is the consequences of the manipulation of the MSS network, and how the avalanches are affected by the \textit{hand-made changes} in the configuration of the system. For example, one may hope that by random relaxing of the MSSs, the spatial extent of the avalanches become more limited and the range of avalanches decrease. This is due to the fact that a MSS can be considered as an agent who, when receives a signal, propagates (scatters) it symmetrically in $2d$ ($d\equiv$ the spatial dimension) directions. Relaxing MSSs means decreasing these agents, and equivalently decreasing the range of the corresponding avalanche. In practice, the sandpiles due to their non-linear structure, show more rich structure with various phases under such a manipulation, and needs a detailed simulation which is the aim of the present paper.
\\
We call a local external relaxation \textit{the fillip}. By performing the fillips in a regular rate within the avalanches we show that the system experiences a phase within which some oscillations occur between two limits. We control the strength of the effect of fillips by a parameter, namely $\zeta$, which is defined as follows: suppose that after $n_1$ topplings, $n_2$ fillips are applied, then $\zeta\equiv \frac{n_2}{n_1}$. We will observe that, things are not as simple that mentioned above, and under applying fillips, the system shows some non-linear unexpected behaviors. Especially a bifurcation takes place in some critical $\zeta^*$, above which the average energy of the system oscillates between two $\zeta$-dependent values. A similar phenomenon is also observed for auto-correlation functions, with some $\zeta$-dependent frequencies.\\
The paper has been organized as follows: In the following section, we motivate this study and introduce and describe the model. The numerical methods and the results are presented in the section~\ref{NUMDet} which contains critical, as well as off-critical results. We end the paper by a conclusion.

\section{The construction of the problem}
\label{sec:model}
\begin{figure}
	\centerline{\includegraphics[scale=.35]{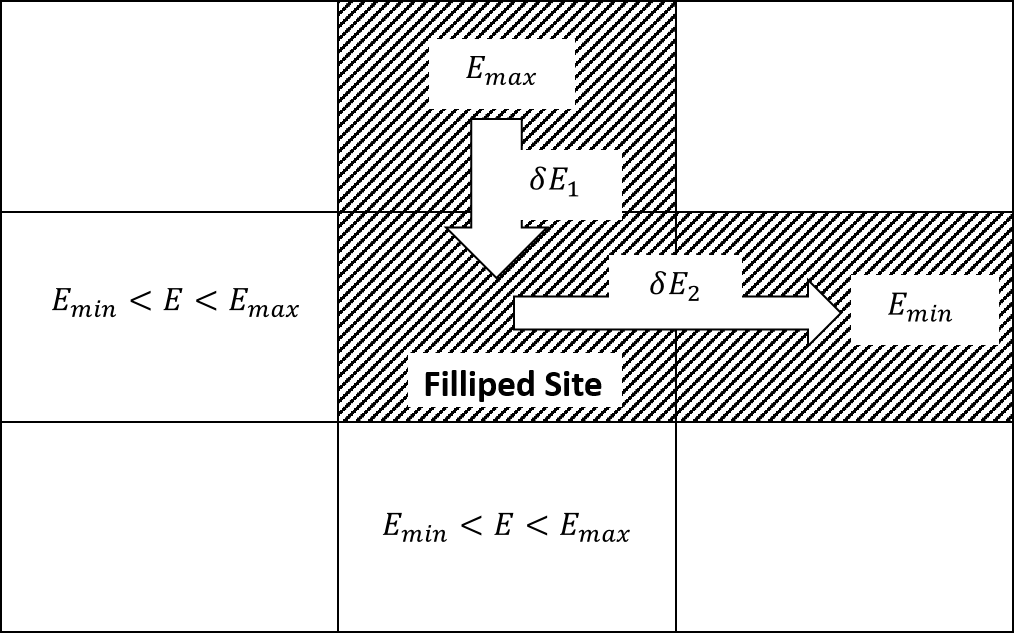}}
	\caption{Schematic of a fillip. The shaded sites are affected ($\delta E_1$ ($\delta E_2$) energy units transfer from (into) the site $i_{\text{max}}$ ($i_{\text{min}}$)) by the fillip. For the rules of the energy transfer ($\delta E_1$ and $\delta E_2$) refer to the text.}
	\label{fig:schematic}
\end{figure}
Let us first define the model in a square $L\times L$ lattice in fillip-less limit, i.e. $\zeta=0$ which is the ordinary (continuous) BTW model. In this model, each site $i$ has an integer height (energy) $ E_i \ge 1$. At the initial state, one can set randomly the energy of each site with the restriction $ E_i \le E_c \equiv 4n$, in which $n$ is an arbitrary integer, and is set to $10$ in this letter. At each time step an energy unit (as a single stimulation) is added to a randomly chosen site ($ E_i \to E_i+1 $). This can be considered as a slow external driving and leads to a fast relaxation process (avalanche) within the system.  This relaxation consists of a conservative redistribution of the energy at sites, i.e. if the energy content of a site exceeds $E_c$, then $ E_i\to E_i+\Delta_{i,j} $ in which $\Delta_{i,j}=-E_c$ if $i=j$, $\Delta_{i,j}=n$ if $i$ and $j$ are neighbors and zero otherwise. A toppling may cause the nearest-neighbor sites to become unstable (have energies larger than $E_c$) and topple in their own turn and so on, until all sites over the lattice are stable.\\
Now let us introduce the fillips. A fillip is defined as the action in which a site is chosen randomly and is checked for a more stable configuration. To do this, the energy content of its neighbors is checked. Suppose that the selected site is $i$ with nearest neighbors $i_1$, $i_2$, $i_3$ and $i_4$. Among these neighbors, label $i_{\text{max}}$ as the site in which $E(i_{\text{max}})=\text{Max} \left\lbrace E_i \right\rbrace_{i=1}^{4} $, and $i_{\text{min}}$ as the site in which $E(i_{\text{min}})=\text{Min} \left\lbrace E_i \right\rbrace_{i=1}^{4}$. A fillip is composed of two updates: $\delta E_1\equiv \text{int}\left[(E(i_{\text{max}})-E(i))/2 \right]$ energy units (if positive) flow from the site $i_{\text{max}}$ to $i$, and then $\delta E_2\equiv \text{int}\left[(E(i)-E(i_{\text{min}}))/2 \right]$ energy units (if positive) flow from the site $i$ to $i_{\text{min}}$. In case of more-than-one sites having the same (maximum or minimum) energy, the site from/into which the energy flows is chosen randomly. If the site $i$ is locally maximum (minimum), automatically no energy flows into (from) the site to the neighbors. No fillip is applied to the unstable sites. Therefore, we have two kinds of relaxations: the sites which are unstable topple and the sites which are chosen for fillips moderate their local gradient of energy. We call the first procedure as the \textit{toppling} and the second one as the \textit{fillip}. This problem can also be called a \textit{diffusive sandpile model}, in which the sand grains are lubricated such that they have the chance to slip to the neighboring sites. The local relaxation in a fillip has been schematically shown in Fig.~\ref{fig:schematic}. \\
Let us consider the problem from another point of view. Applying local relaxations (fillips) means that the system is being smoothed and the number of MSSs decreases, i.e. the avalanches become lower in range, and the toppling events in the avalanches decrease. As a result, the number of energy units that leave the system decrease, and accordingly, the balance between external drive and the dissipated energy is displaced. This cannot last for much time, since the average energy of the system, which grows with external drive, cannot become larger than $E_{th}$. In the other words, when $\bar{E}\rightarrow E_{th}$, under external drives some very large avalanches take place. This can be understood noting that in this limit the number of MSSs is nearly equal to $L^2$. Let us consider the problem in the mean field (MF) level. The average energy at time $T$ ($T$th injection) is considered to be $\bar{E}(T)$. Then one readily finds that, on average $\bar{E}(T+1)=\bar{E}(T)+\frac{1}{N}-4L\frac{A(T)}{N}$ in which $N=L^2$ is the number of sites in the system, and $A(T)$ is the avalanche size in time $T$. For the steady state (in which the average energy is nearly $T$-independent), one finds $A(T)=\frac{1}{4L}$. Now we consider the effect of fillips by decreasing the avalanche sizes by a $\zeta$-dependent scale factor; $A'(T)=f(\zeta)A(T)$, which leads to $\delta\bar{E}(T)\equiv \bar{E}_{T+1}-\bar{E}_{T}=\frac{1}{N}\left(1-f(\zeta)\right)$. But for the case in which $A'(T)\approx L^2$ (or equivalently $\bar{E}(T)\rightarrow E_{th}$), $\delta\bar{E}=-\left(4L-1\right)$, leading to the following equation for the conditional probability:
\begin{equation}
P(\bar{E}_{T+1}=z|\bar{E}_T=M) = \left\lbrace \begin{array}{ll}
\delta_{z,M+\frac{1}{N}(1-f(\zeta))} & M<E_{th}\\
\delta_{z,M-4L+1} & M\approx E_{th}
\end{array}\right. 
\end{equation}
Using this function, one can calculate the branching ratio, defined by $b(M)\equiv\textbf{E}\left[\frac{\bar{E}_{T+1}}{M}|\bar{E}_T=M \right]$, to be:
\begin{equation}
b(M)=\left\lbrace \begin{array}{ll}
1+\frac{1-f(\zeta)}{NM} & M<E_{th} \\ 1-\frac{4L-1}{M} & M\approx E_{th}
\end{array}\right. 
\label{Eq:meanfield}
\end{equation}
in which $\textbf{E}\left[ \ \right]$ is the expectation value. Note that when $b(M)>1$ ($b(M)<1$), then for a given $M$, the average energy of the system will increase (here linearly) (decrease, here abruptly) with $T$. This relation predicts that a bifurcation takes place at a non-zero $\zeta$, above which some oscillations occur. For the first branch, the mean energy increases linearly with $T$ up to time at which $\bar{E}\approx E_{th}$. At this point the average energy drops abruptly by $\delta \bar{E}\approx -4L$ (corresponding to the lower branch). This is accompanied with some large avalanches, which are named as \textit{spanning avalanches (SA)}. This dropping should be independent of $\zeta$.\\

\subsection{Numerical details}

\begin{figure*}
	\centering
	\begin{subfigure}{0.3\textwidth}\includegraphics[width=\textwidth]{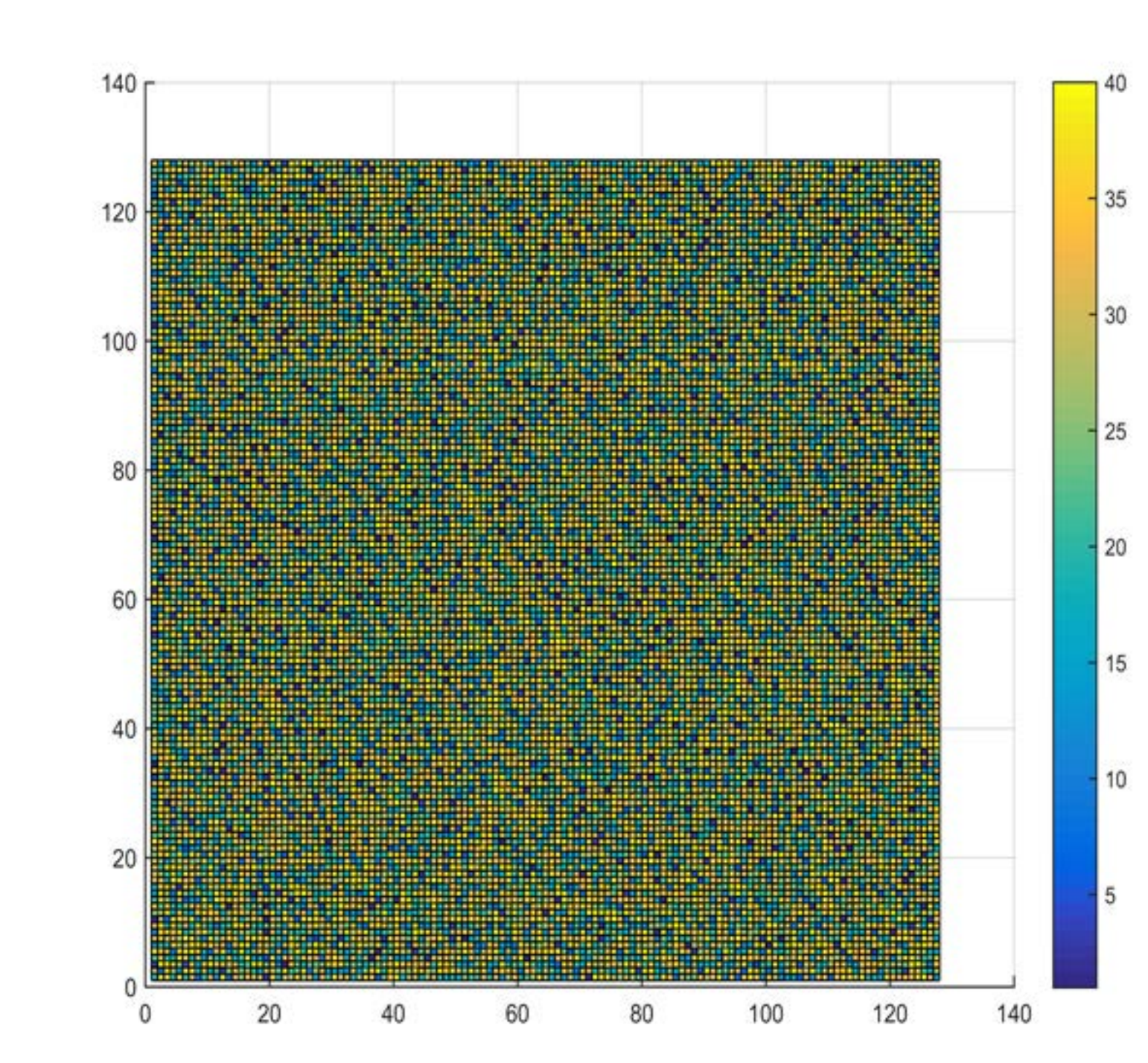}
		\caption{}
		\label{fig:m0}
	\end{subfigure}
	\begin{subfigure}{0.3\textwidth}\includegraphics[width=\textwidth]{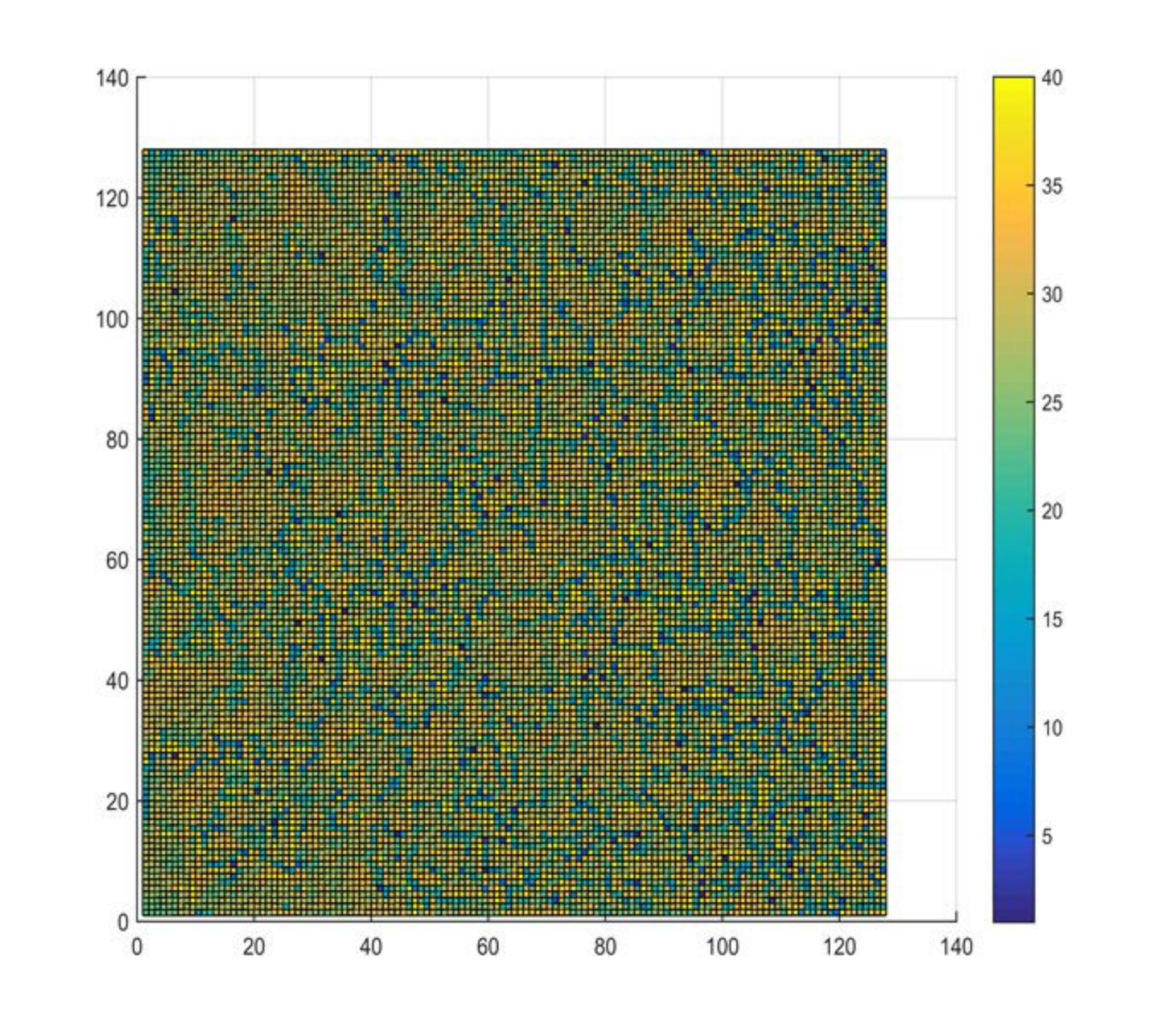}
		\caption{}
		\label{fig:m0001}
	\end{subfigure}
	\begin{subfigure}{0.3\textwidth}\includegraphics[width=\textwidth]{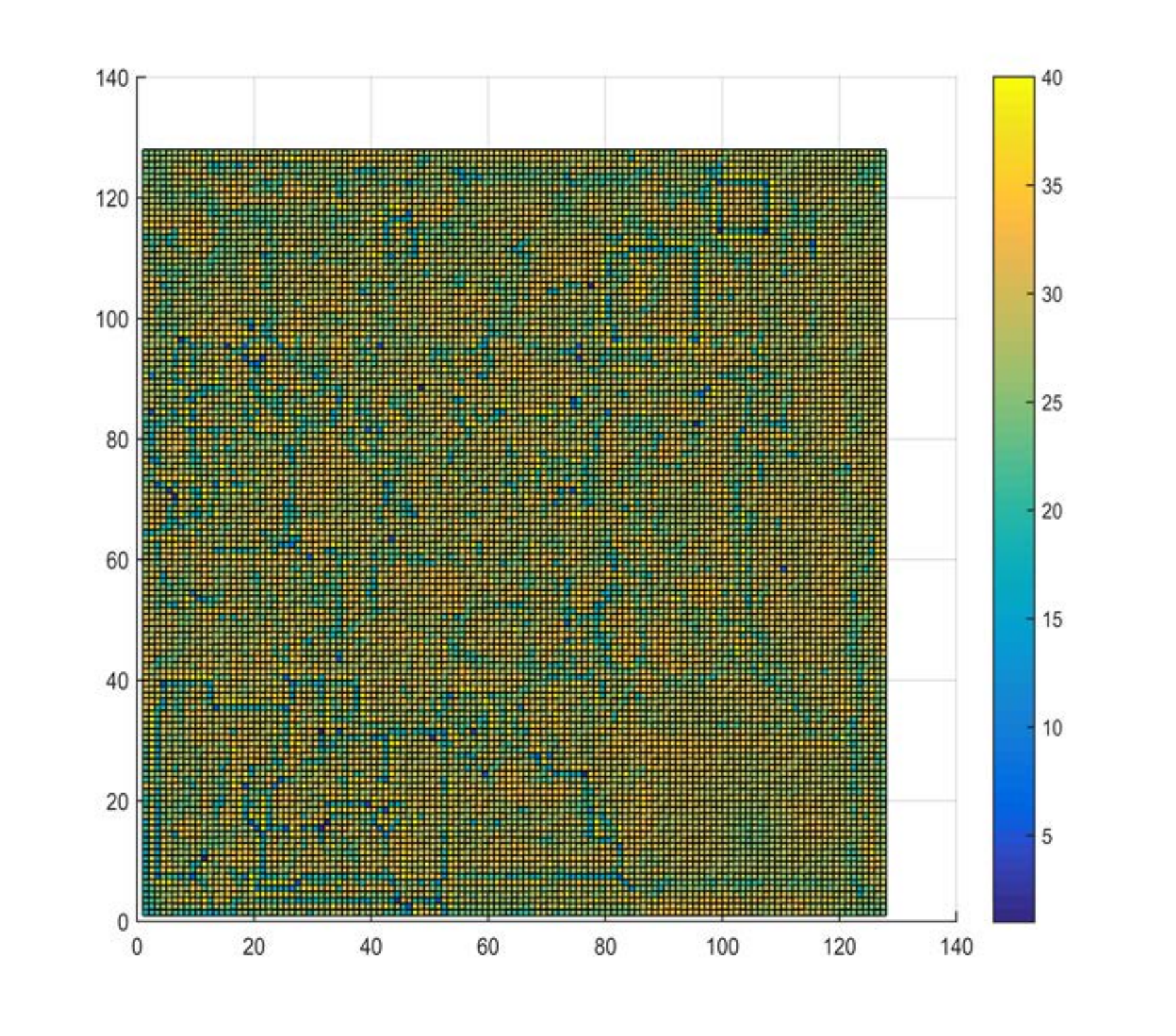}
		\caption{}
		\label{fig:m0005}
	\end{subfigure}
	\begin{subfigure}{0.3\textwidth}\includegraphics[width=\textwidth]{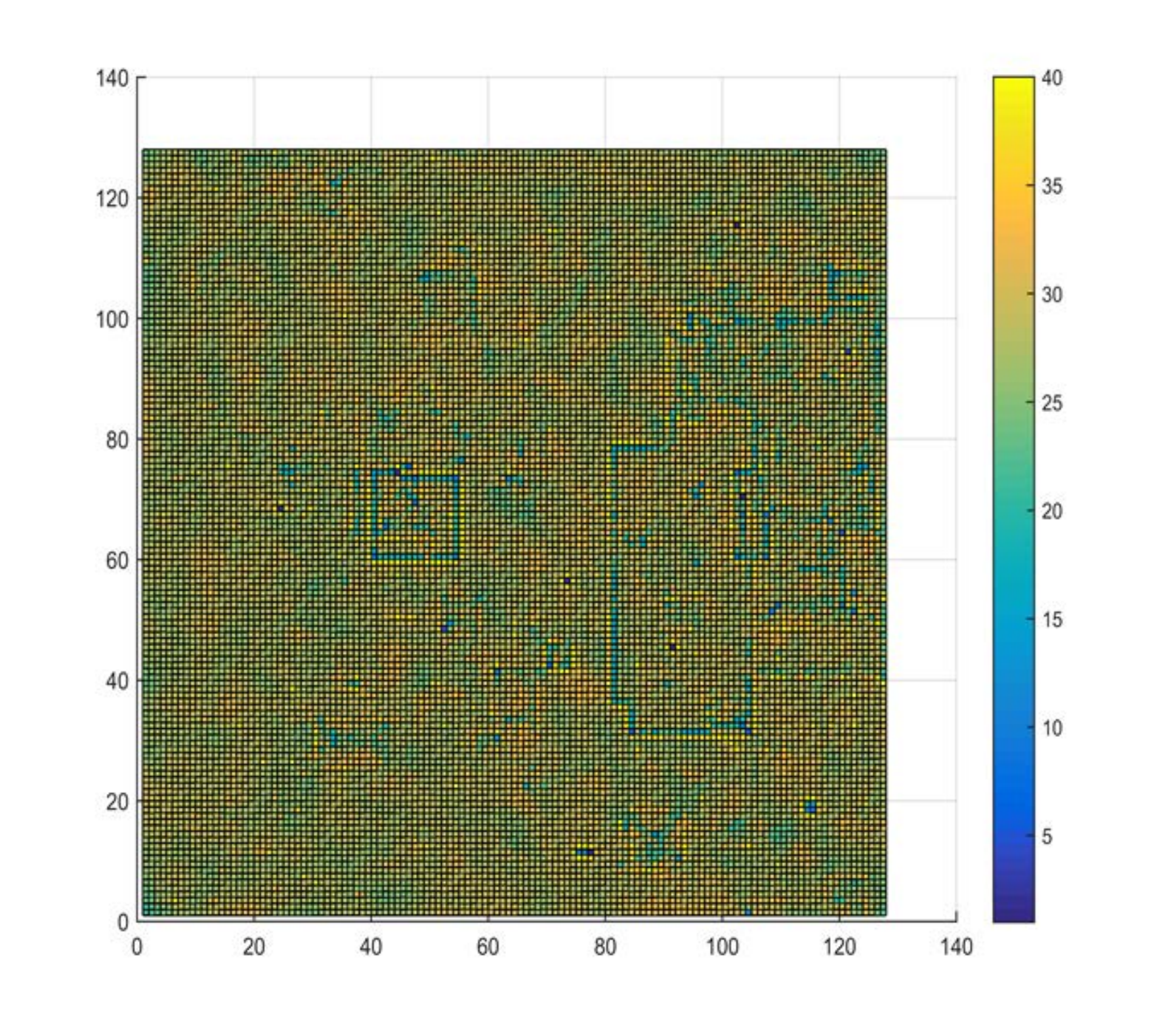}
		\caption{}
		\label{fig:m001}
	\end{subfigure}
	\begin{subfigure}{0.3\textwidth}\includegraphics[width=\textwidth]{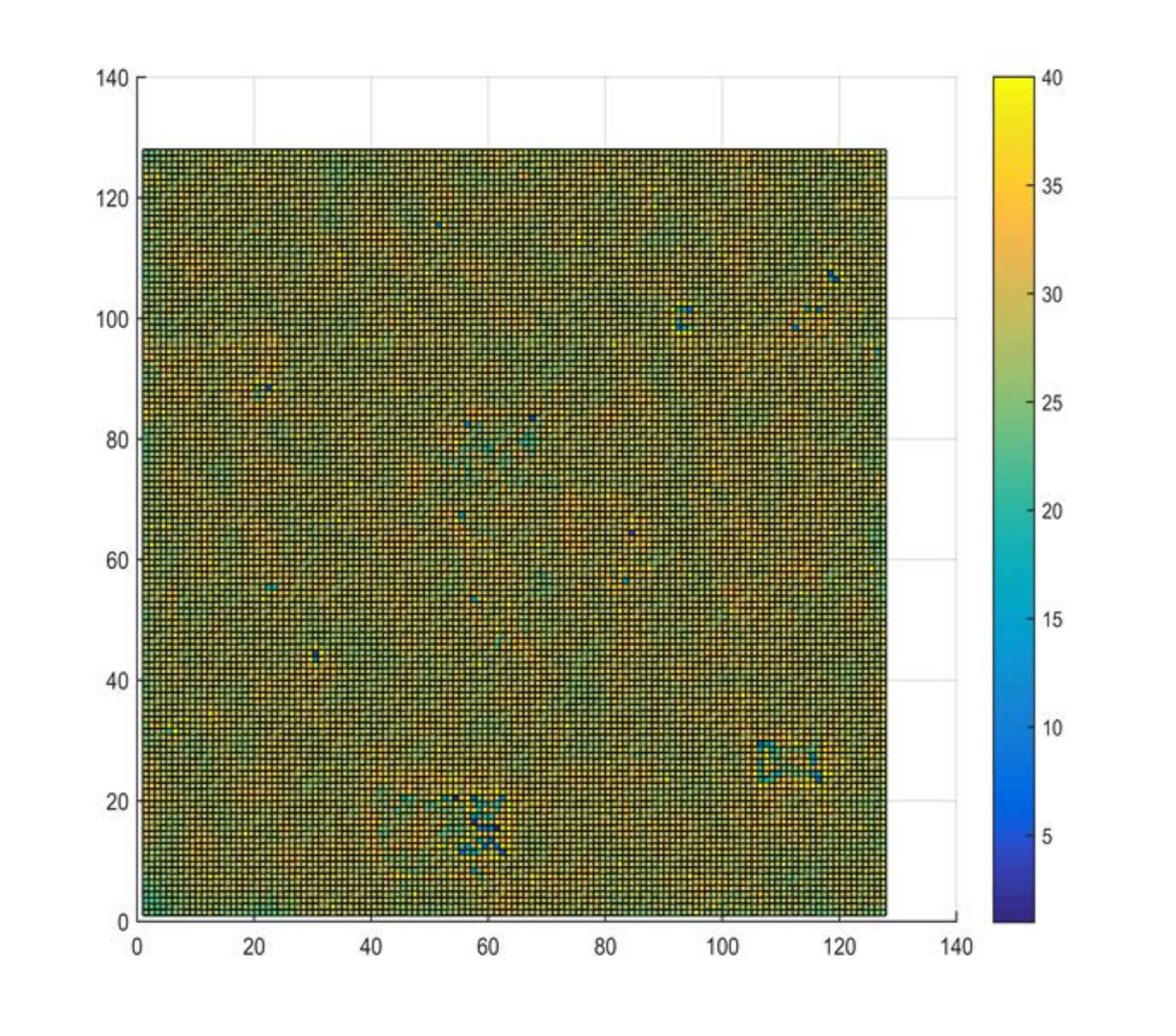}
		\caption{}
		\label{fig:m005}
	\end{subfigure}
	\begin{subfigure}{0.3\textwidth}\includegraphics[width=\textwidth]{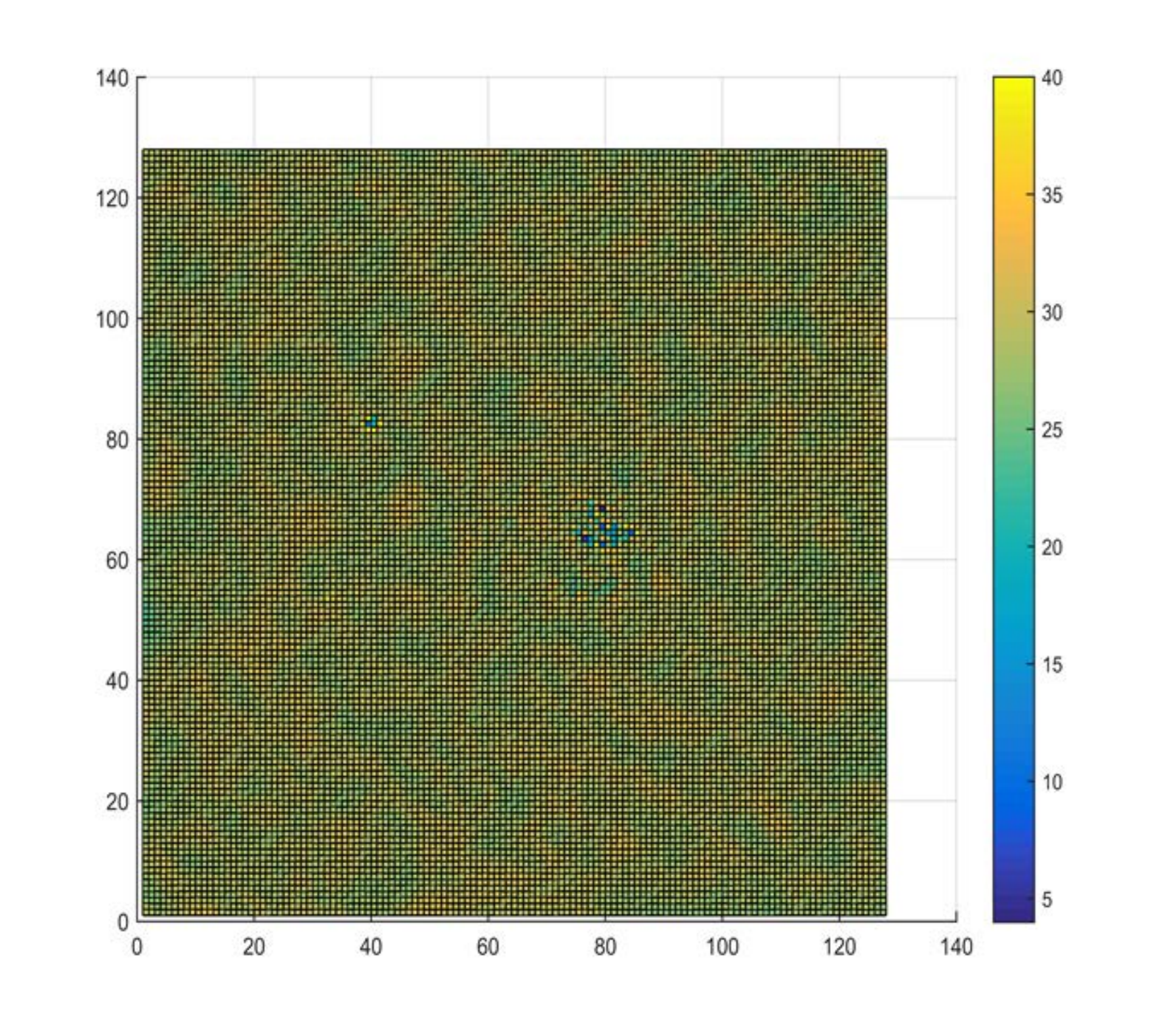}
		\caption{}
		\label{fig:m05}
	\end{subfigure}
	\begin{subfigure}{0.3\textwidth}\includegraphics[width=\textwidth]{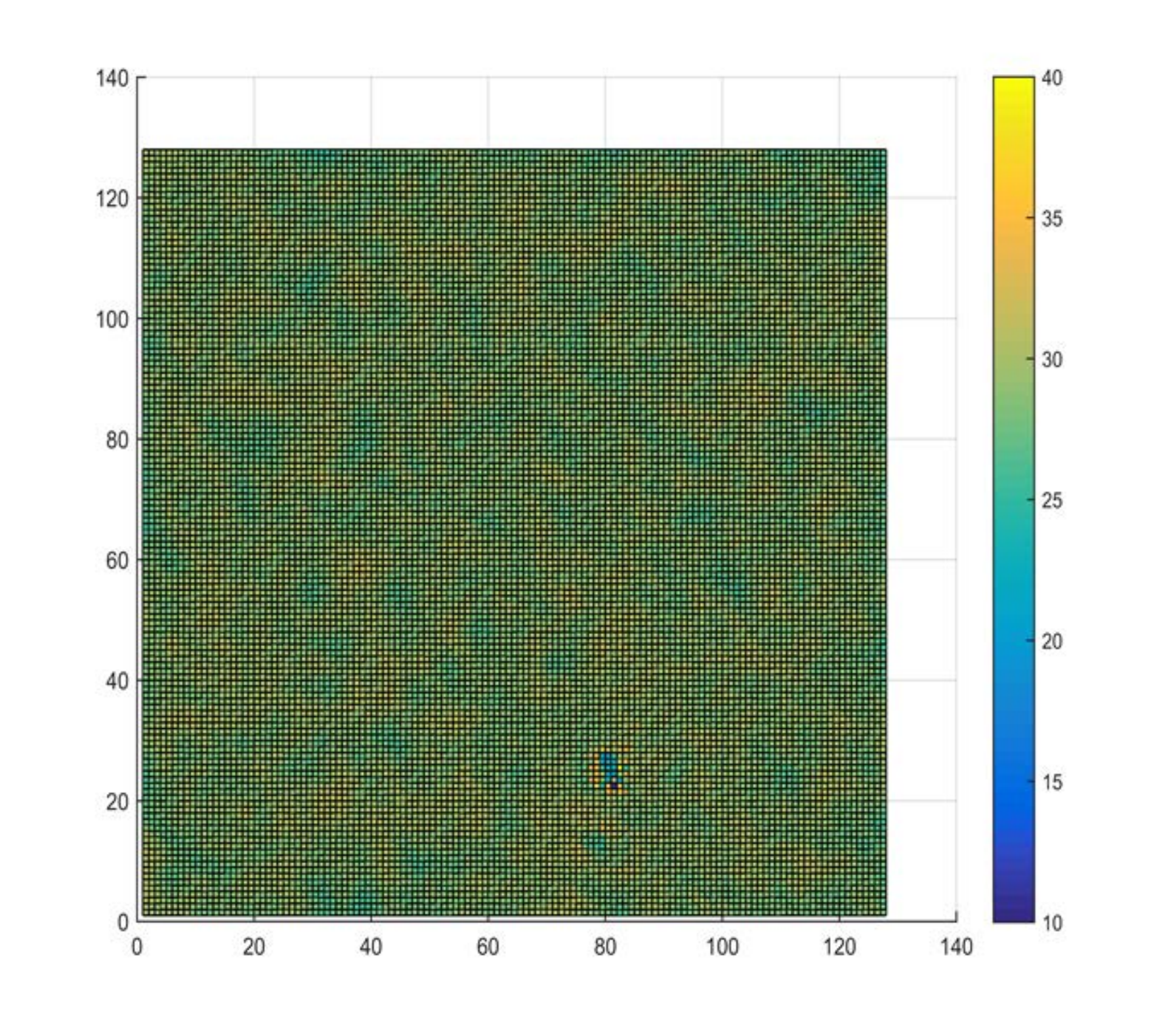}
		\caption{}
		\label{fig:m1}
	\end{subfigure}
		\begin{subfigure}{0.3\textwidth}\includegraphics[width=\textwidth]{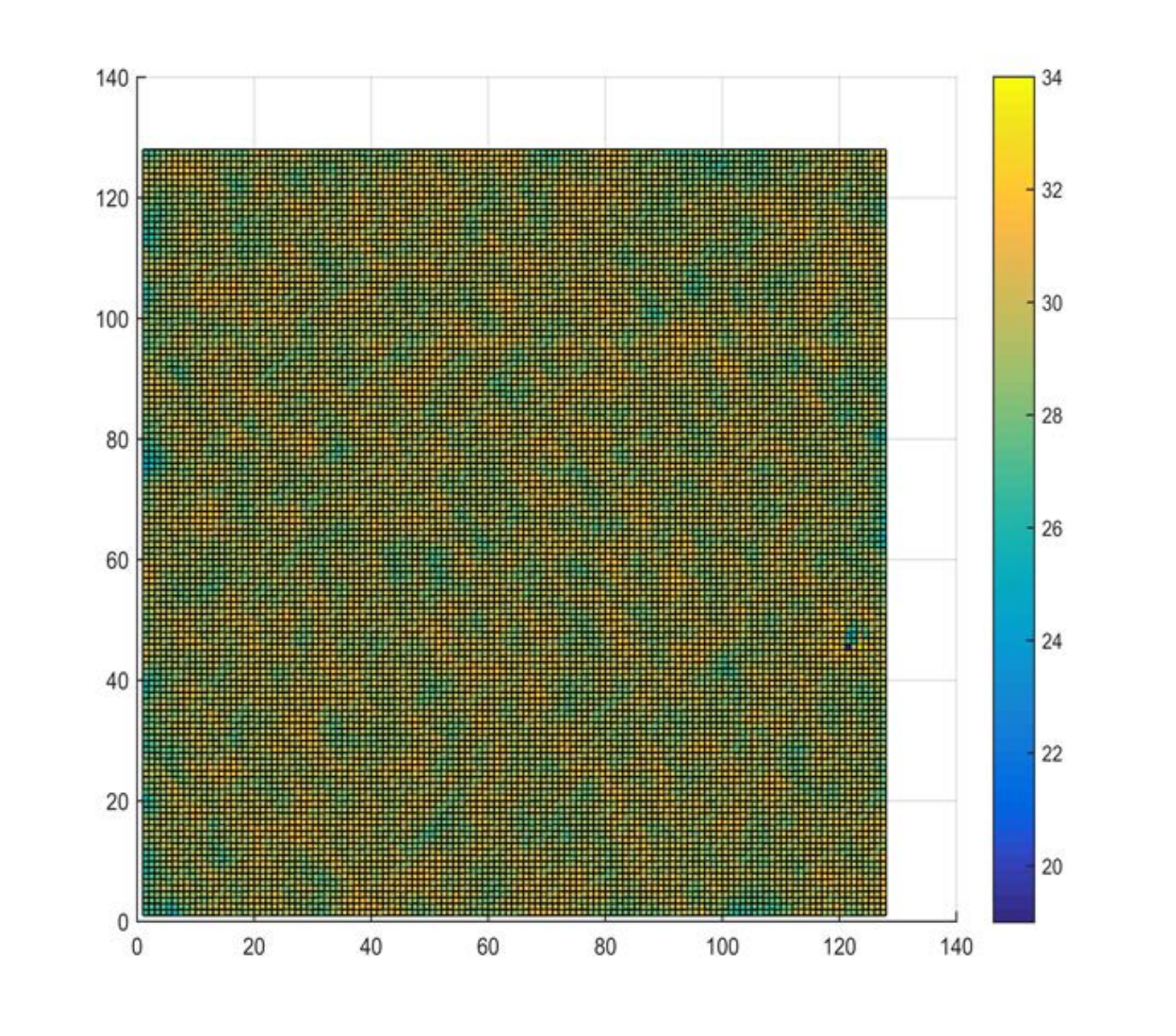}
		\caption{}
		\label{fig:m5}
	\end{subfigure}
		\begin{subfigure}{0.3\textwidth}\includegraphics[width=\textwidth]{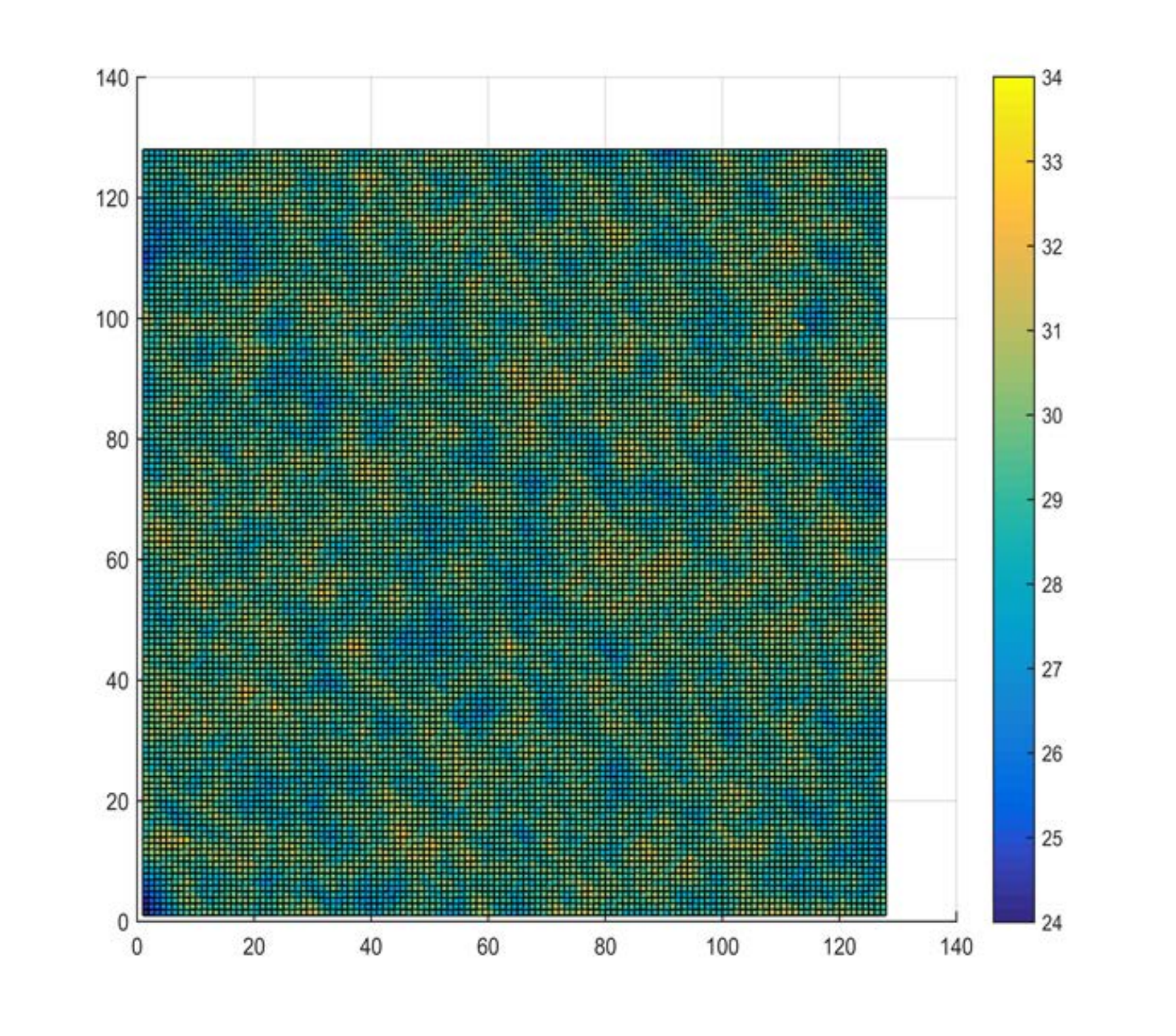}
		\caption{}
		\label{fig:m10}
	\end{subfigure}
	\caption{(Color online): Filliped samples $128\times 128$ with (a) $\zeta=0$ (a) $\zeta=4$, (a) $\zeta=8$, (a) $\zeta=10$, (a) $\zeta=16$, (a) $\zeta=24$, (a) $\zeta=32$, (a) $\zeta=48$, (a) $\zeta=64$.}
	\label{fig:samples1}
\end{figure*}

\begin{figure*}
	\centering
	\begin{subfigure}{0.3\textwidth}\includegraphics[width=\textwidth]{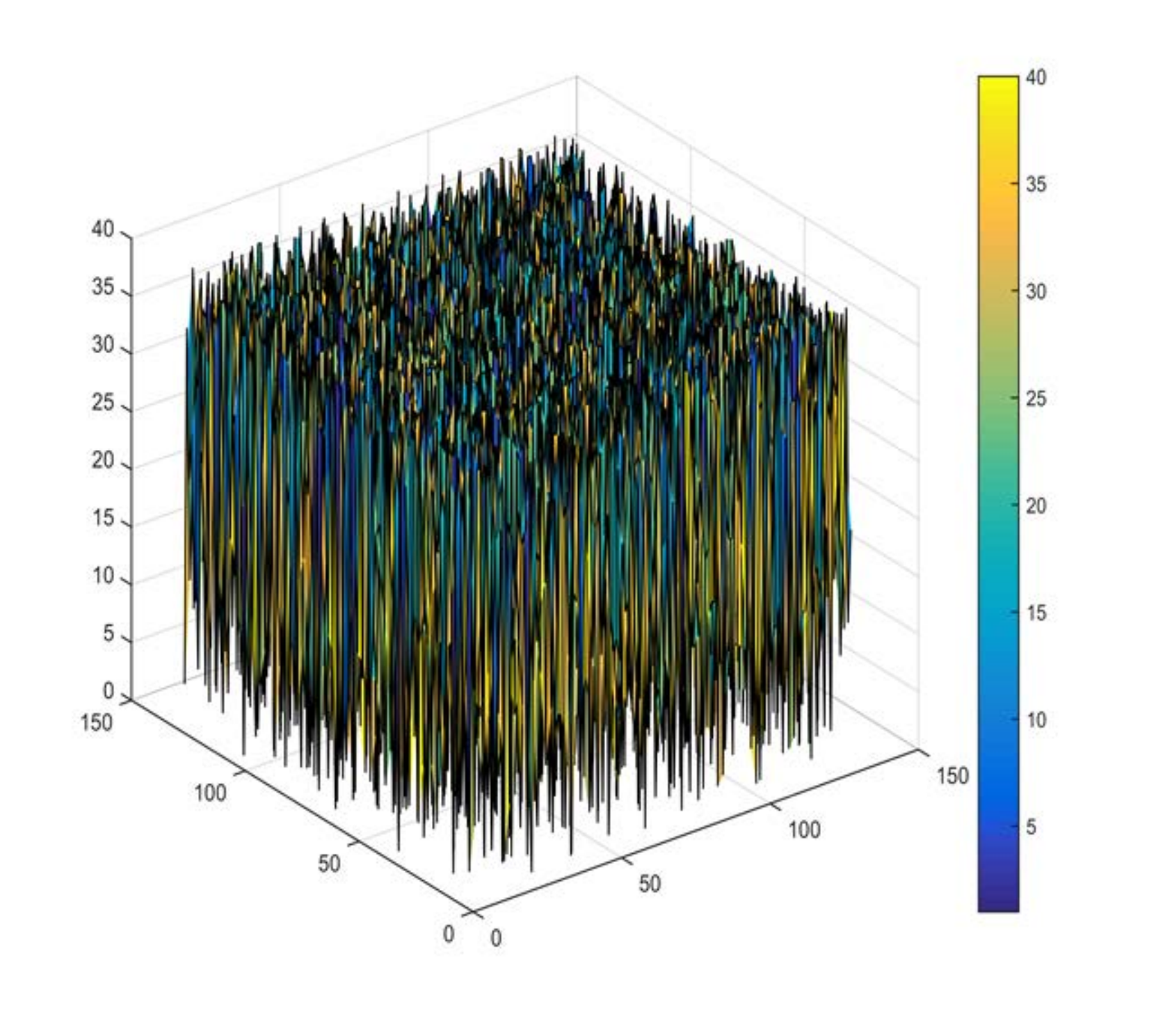}
		\caption{}
		\label{fig:mm0}
	\end{subfigure}
	\begin{subfigure}{0.3\textwidth}\includegraphics[width=\textwidth]{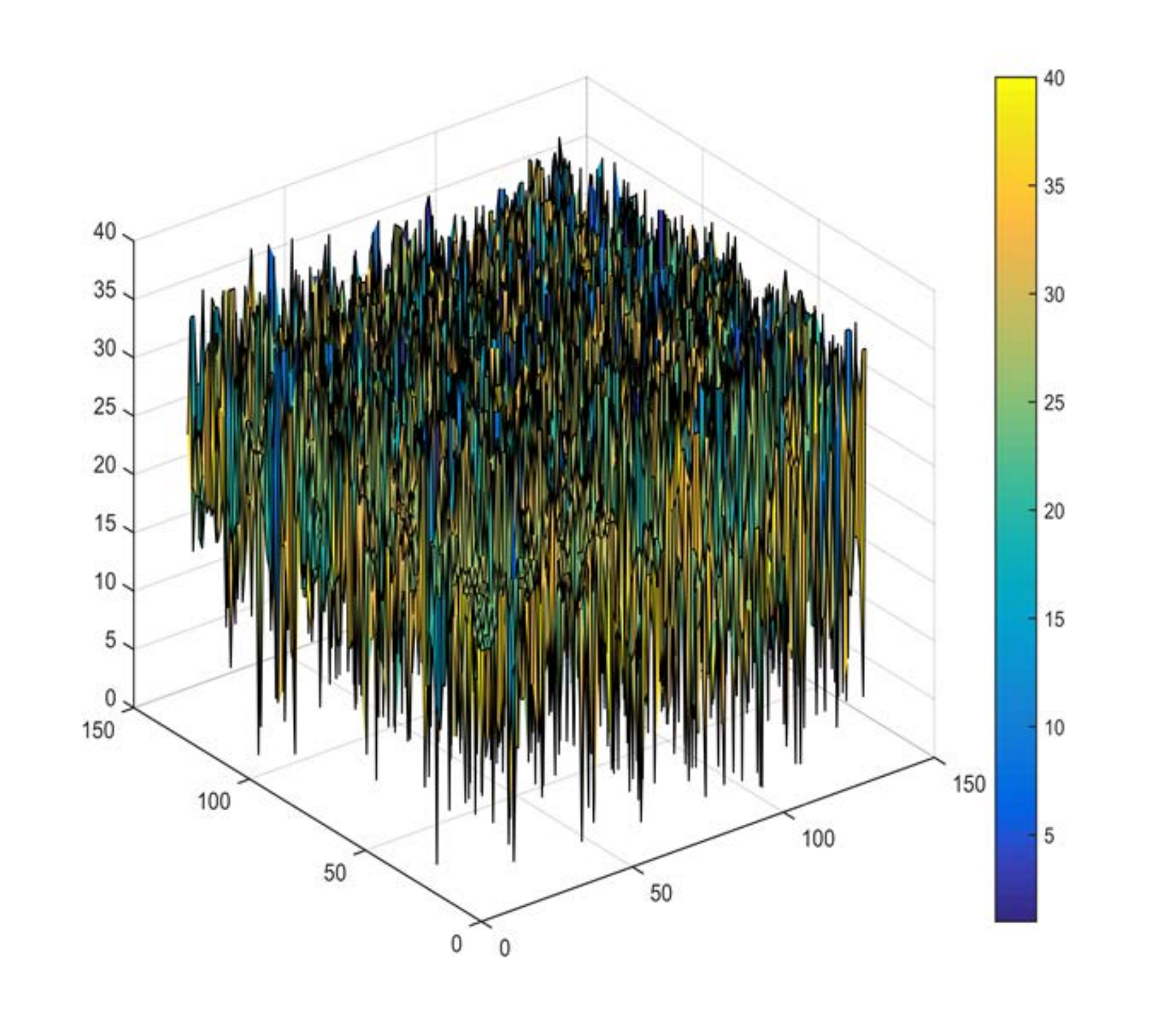}
		\caption{}
		\label{fig:mm0001}
	\end{subfigure}
	\begin{subfigure}{0.3\textwidth}\includegraphics[width=\textwidth]{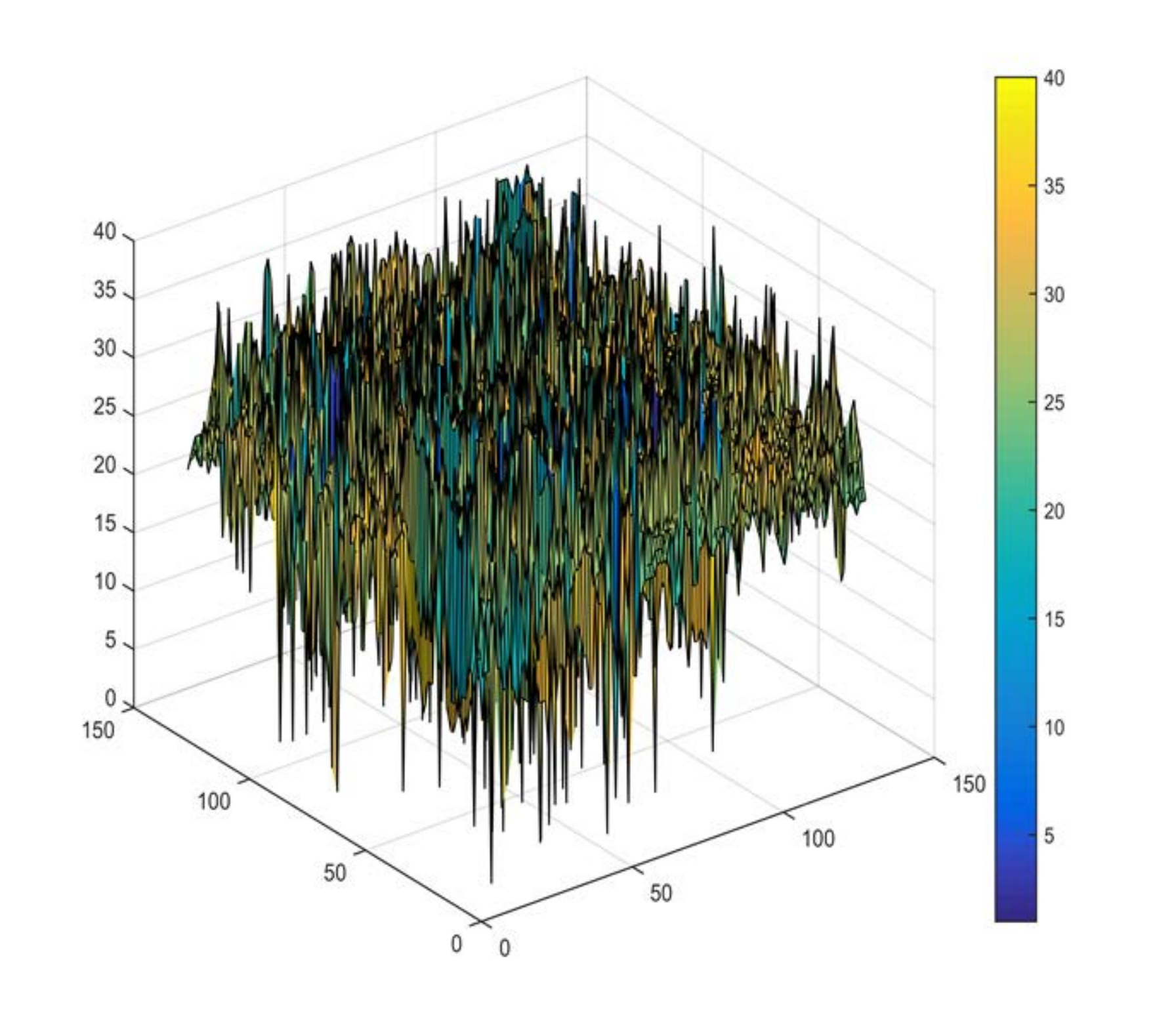}
		\caption{}
		\label{fig:mm0005}
	\end{subfigure}
	\begin{subfigure}{0.3\textwidth}\includegraphics[width=\textwidth]{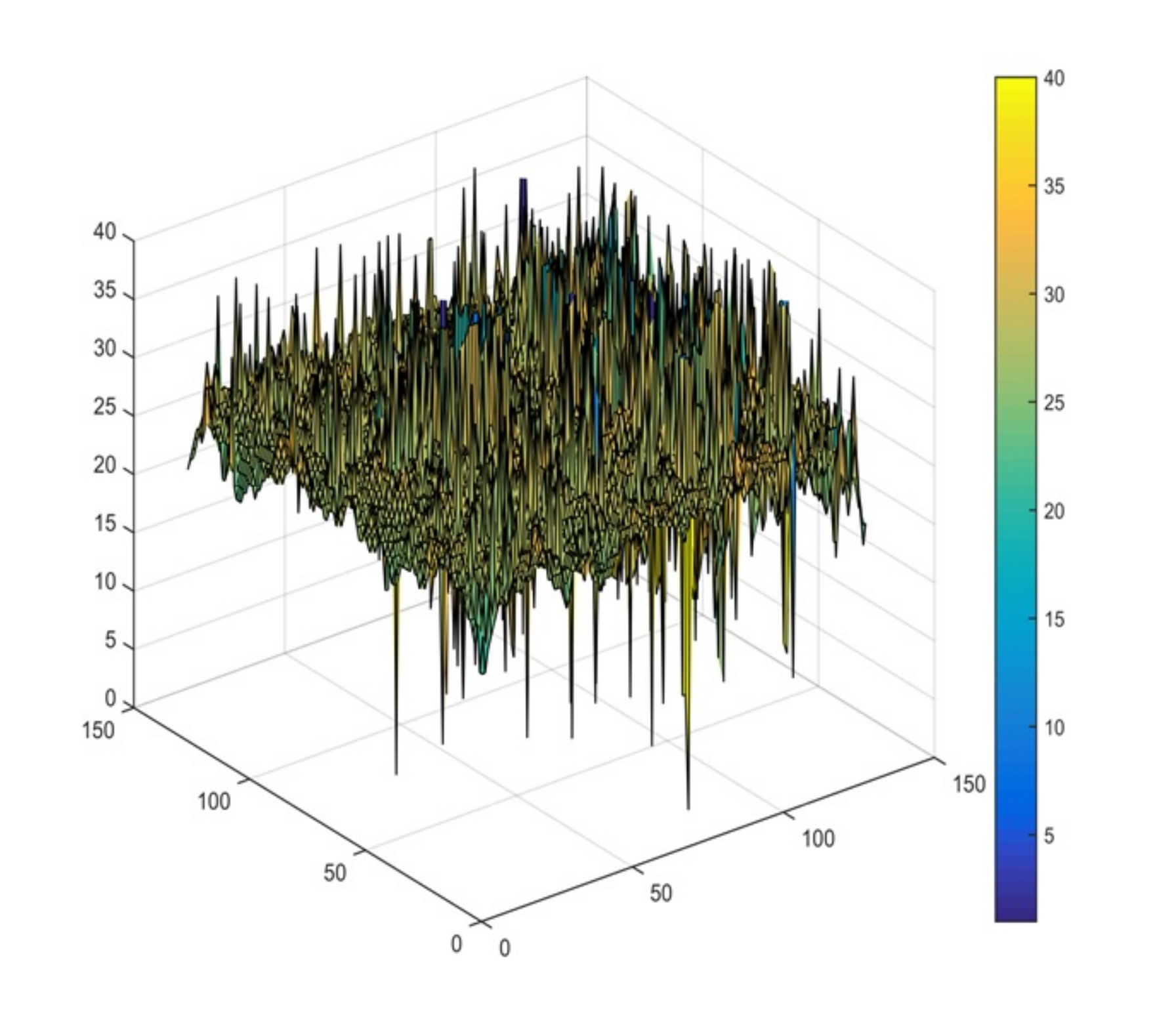}
		\caption{}
		\label{fig:mm001}
	\end{subfigure}
	\begin{subfigure}{0.3\textwidth}\includegraphics[width=\textwidth]{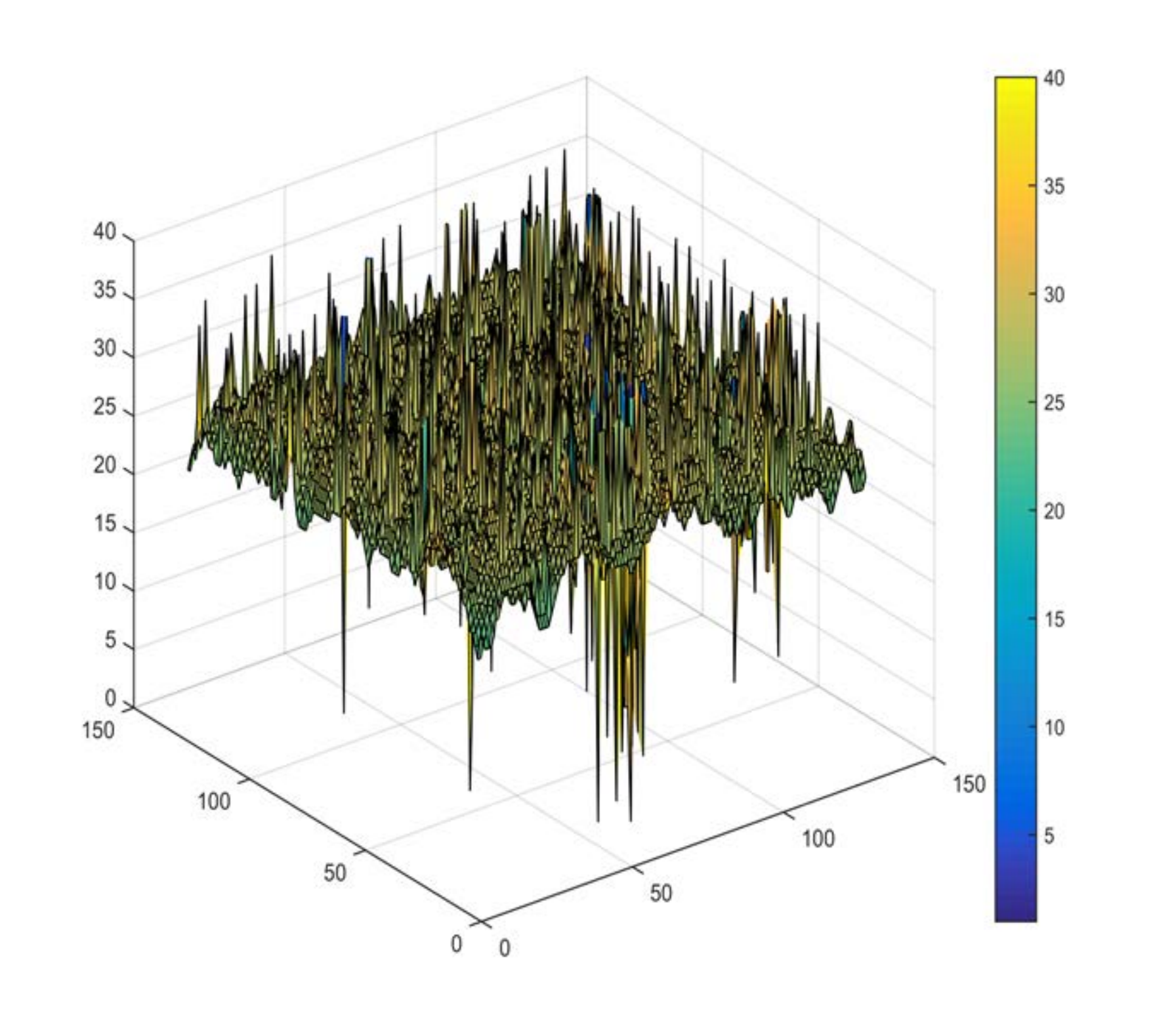}
		\caption{}
		\label{fig:mm005}
	\end{subfigure}
	\begin{subfigure}{0.3\textwidth}\includegraphics[width=\textwidth]{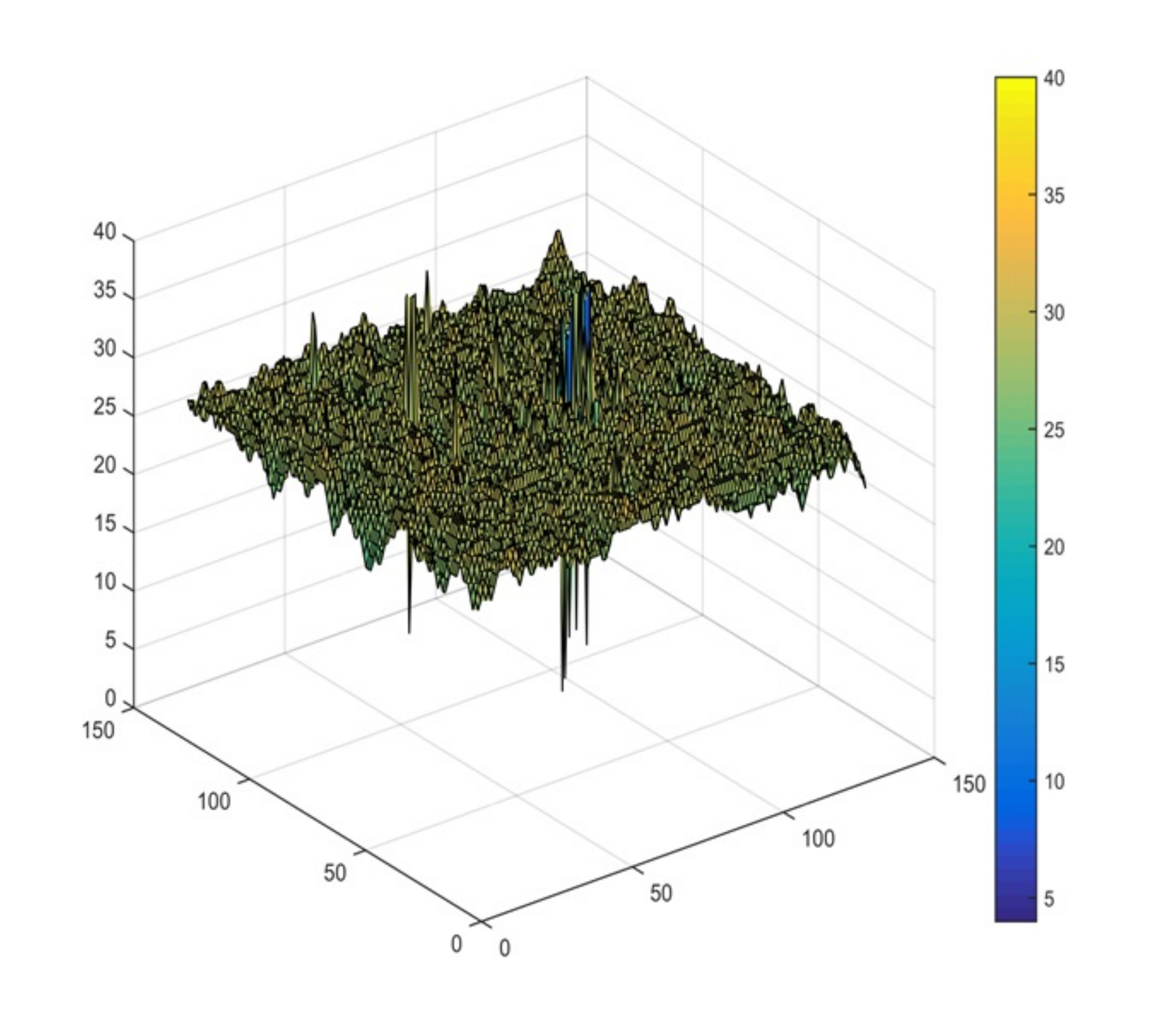}
		\caption{}
		\label{fig:mm05}
	\end{subfigure}
	\begin{subfigure}{0.3\textwidth}\includegraphics[width=\textwidth]{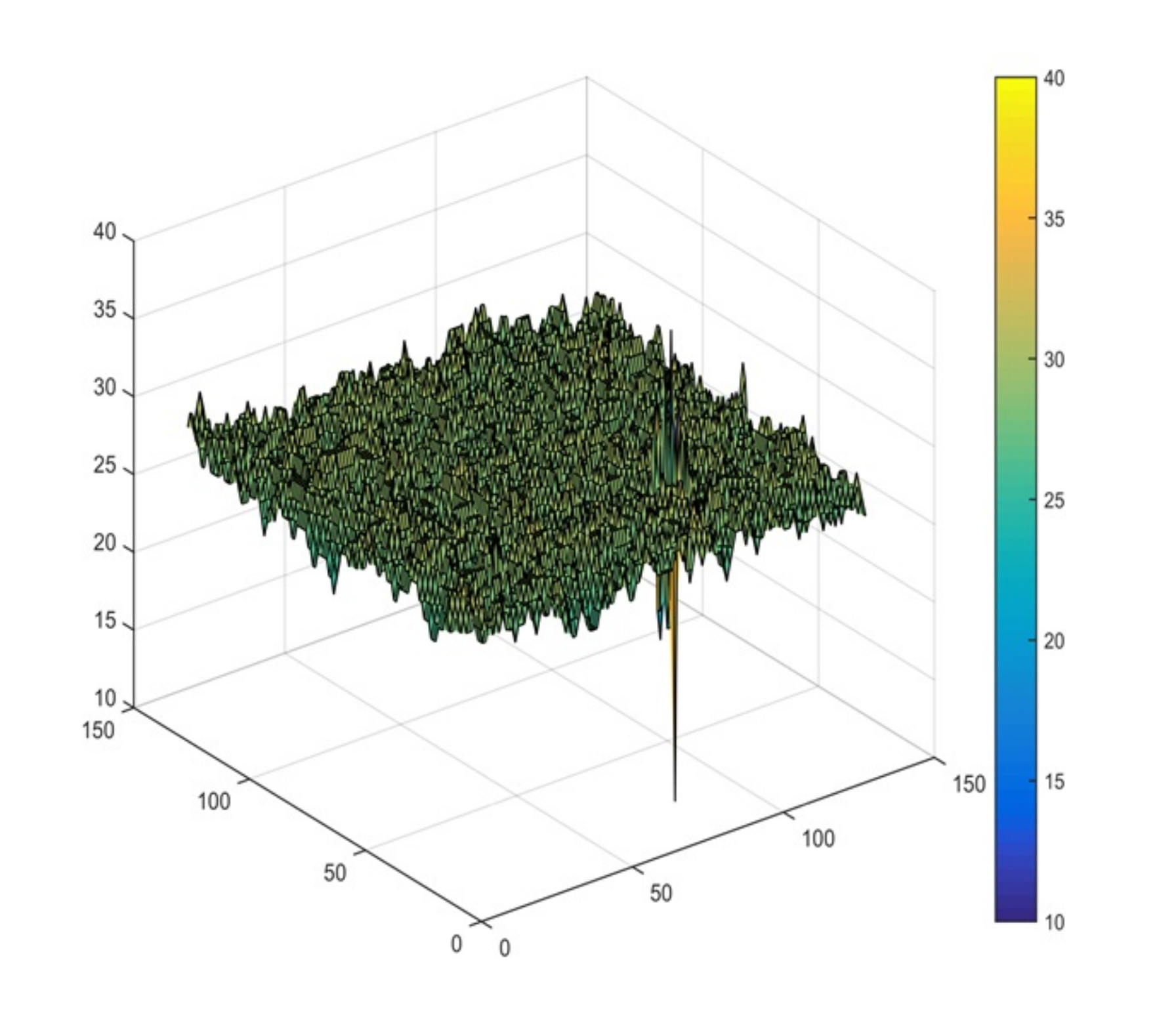}
		\caption{}
		\label{fig:mm1}
	\end{subfigure}
	\begin{subfigure}{0.3\textwidth}\includegraphics[width=\textwidth]{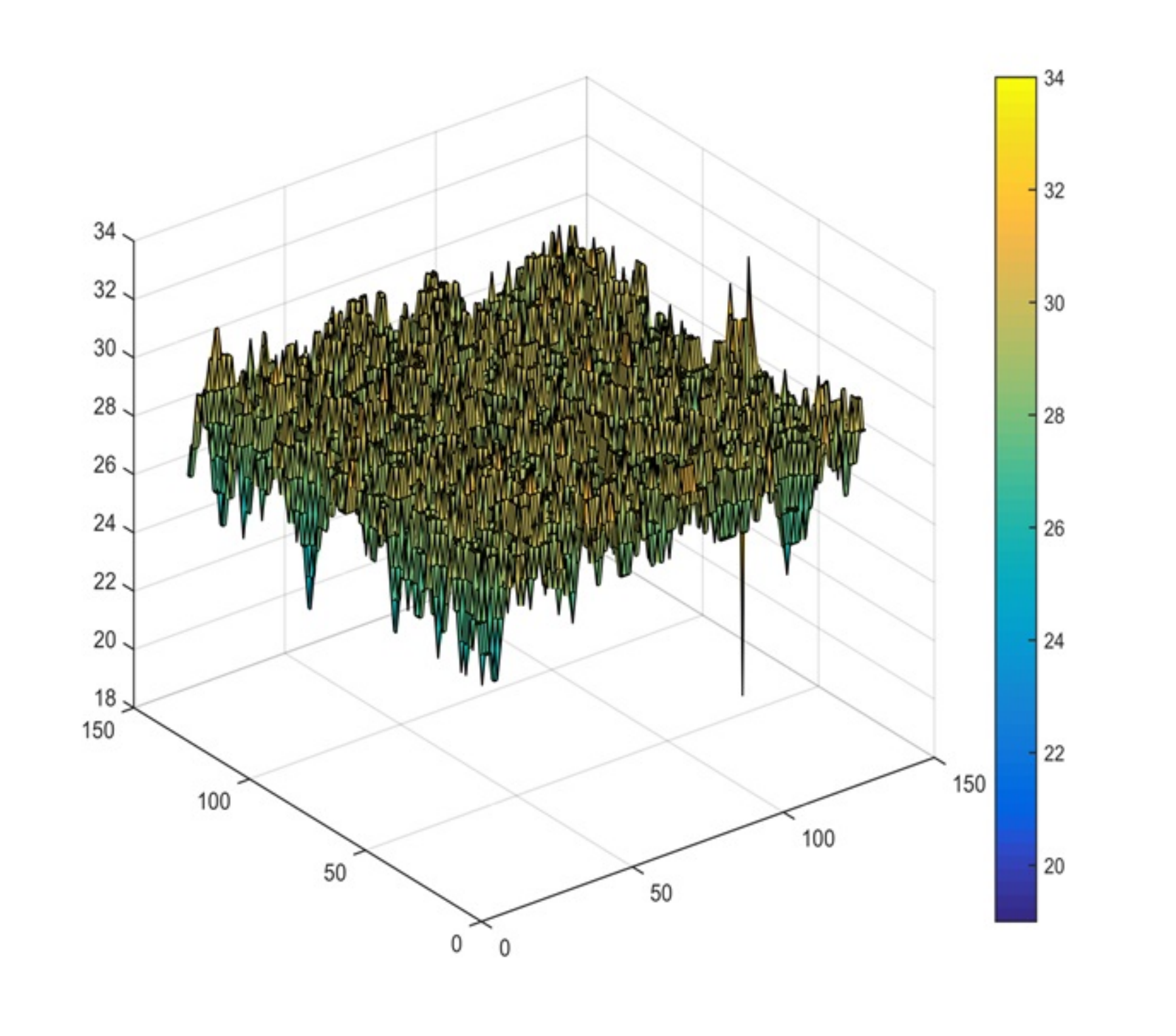}
		\caption{}
		\label{fig:mm5}
	\end{subfigure}
	\begin{subfigure}{0.3\textwidth}\includegraphics[width=\textwidth]{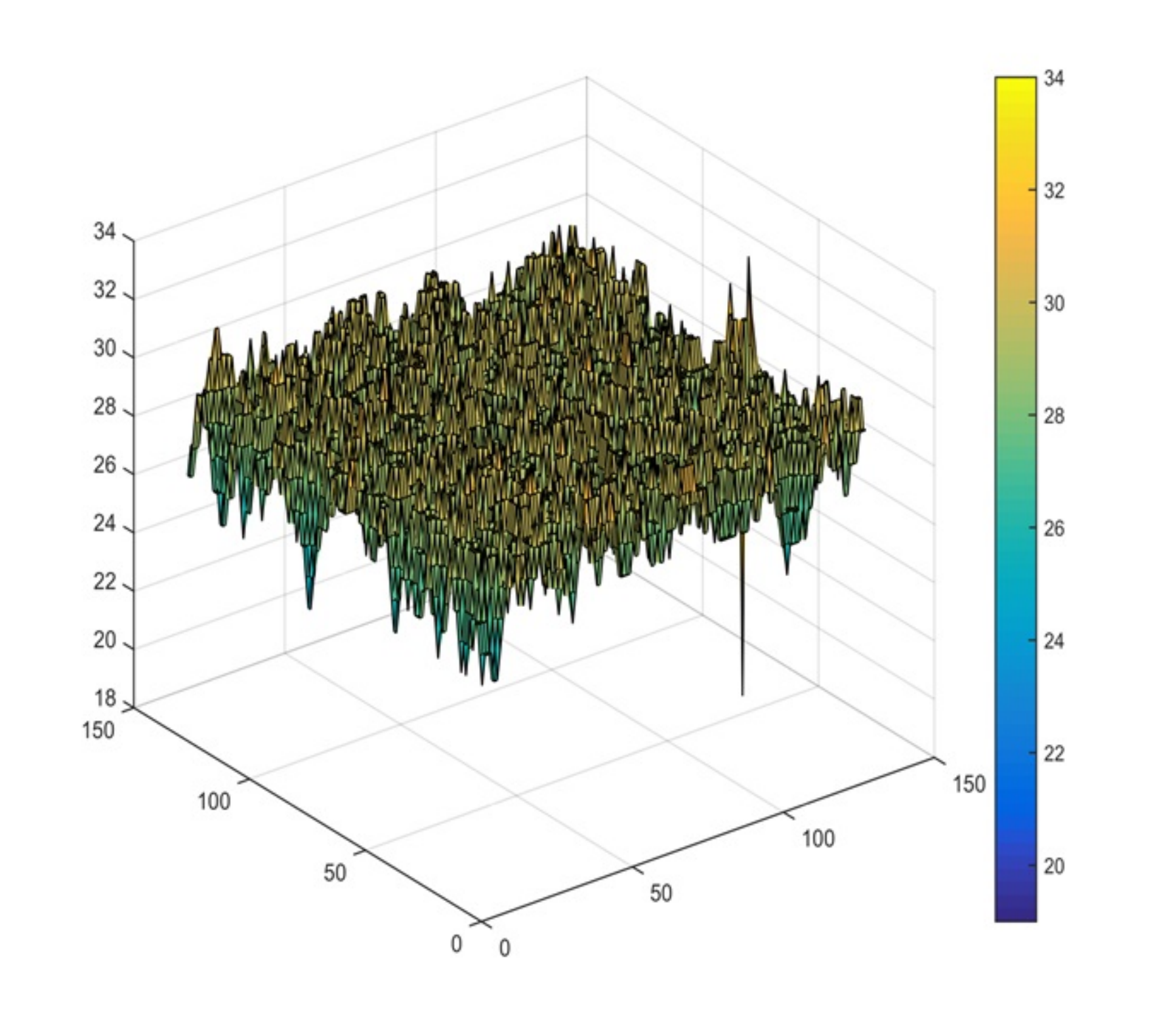}
		\caption{}
		\label{fig:mm10}
	\end{subfigure}
	\caption{(Color online): Roughness of Filliped samples $128\times 128$ with (a) $\zeta=0$ (a) $\zeta=4$, (a) $\zeta=8$, (a) $\zeta=10$, (a) $\zeta=16$, (a) $\zeta=24$, (a) $\zeta=32$, (a) $\zeta=48$, (a) $\zeta=64$.}
	\label{fig:samples2}
\end{figure*}

We have considered $L\times L$ square systems $L=32, 64, 128, 256$ and $512$. The algorithm that has been completely identified in the previous section was employed to extract the samples. We have run the program for various rates of $\zeta$. The run time for small $\zeta$ values was ordinary (like the fillip-less case), and for larger $\zeta$ values, it grew rapidly. For $\zeta=64$, the program was run $100$ days by a $3.2$ GHz CPU. For each $\zeta$ and $L$ over $2\times 10^6$ samples were generated. Some samples have been shown in Figs~\ref{fig:samples1} and~\ref{fig:samples2}.

\section{measures and results}\label{NUMDet}
\begin{figure*}
	\centering
	\begin{subfigure}{0.49\textwidth}\includegraphics[width=\textwidth]{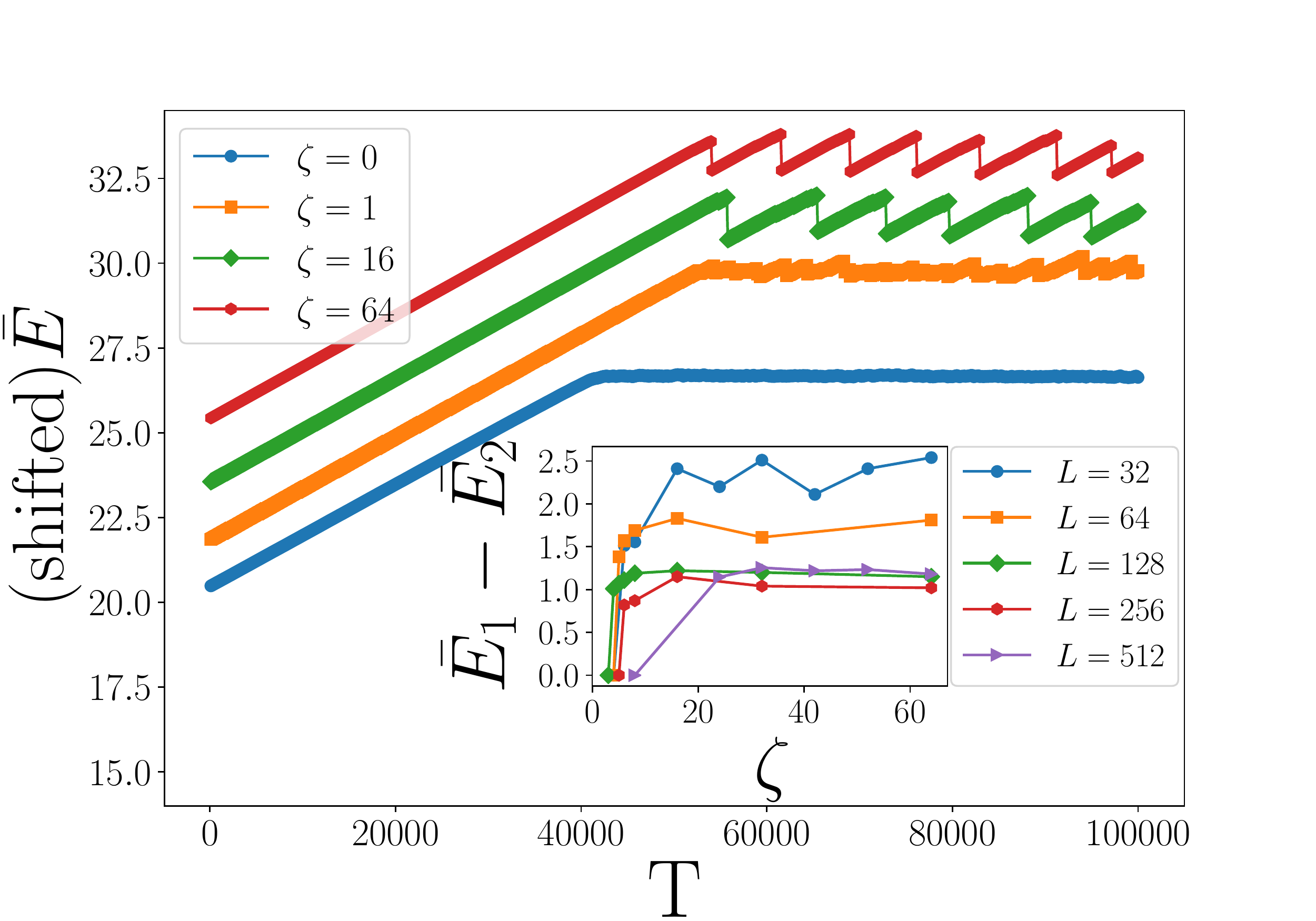}
		\caption{}
		\label{fig:a_ave}
	\end{subfigure}
	\begin{subfigure}{0.49\textwidth}\includegraphics[width=\textwidth]{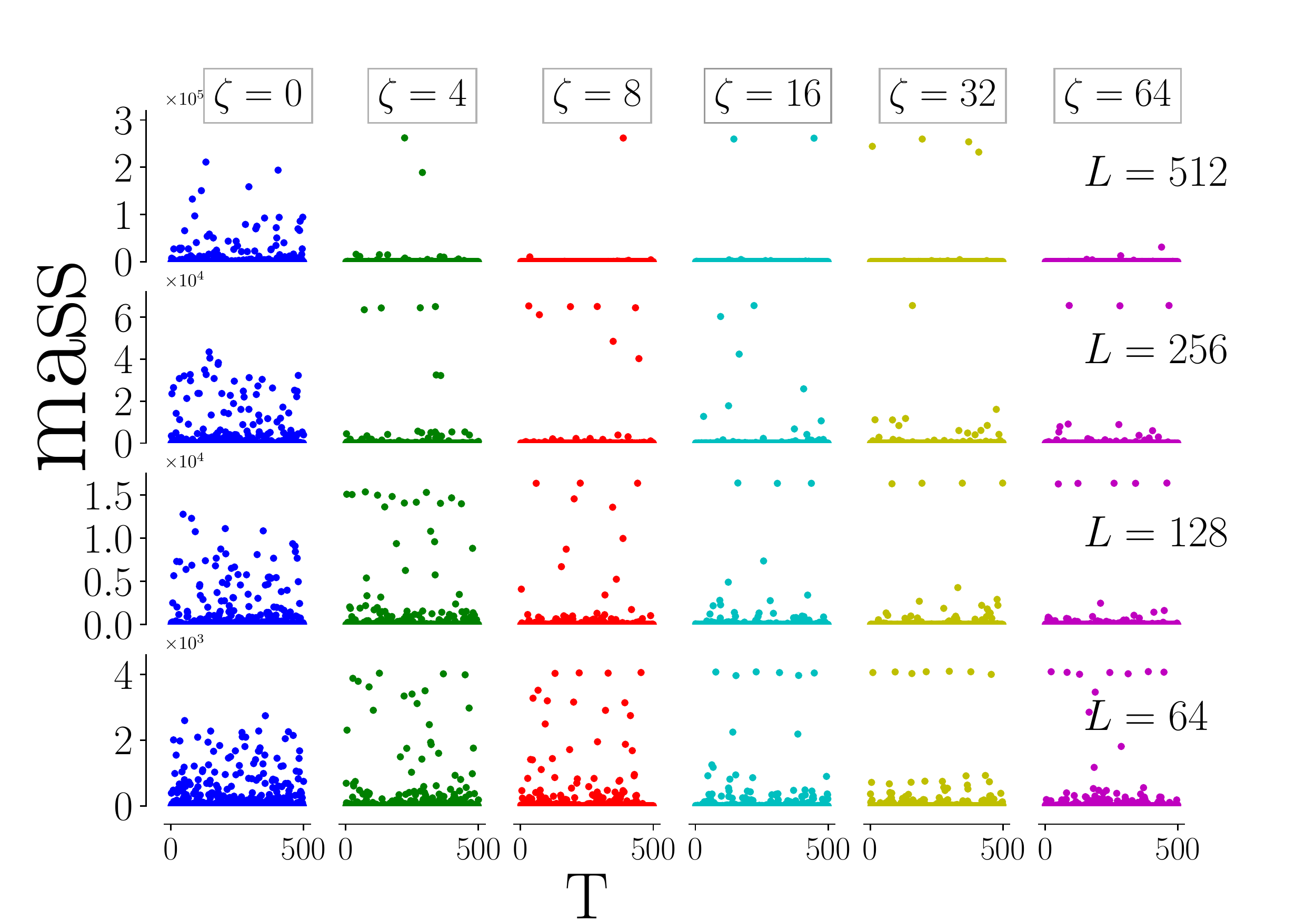}
		\caption{}
		\label{fig:mass_T}
	\end{subfigure}
	\begin{subfigure}{0.49\textwidth}\includegraphics[width=\textwidth]{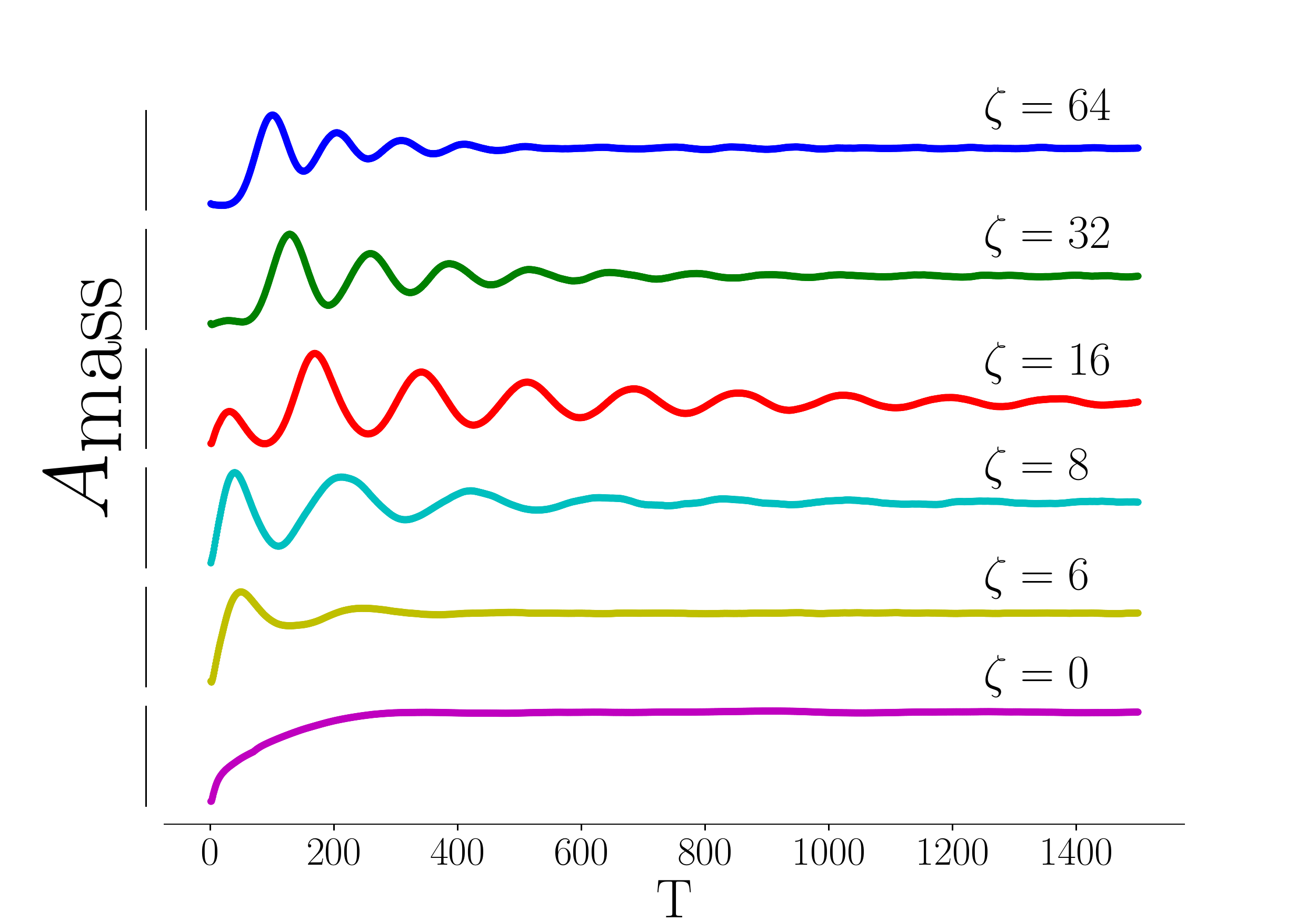}
		\caption{}
		\label{fig:A_mass}
	\end{subfigure}
		\begin{subfigure}{0.49\textwidth}\includegraphics[width=\textwidth]{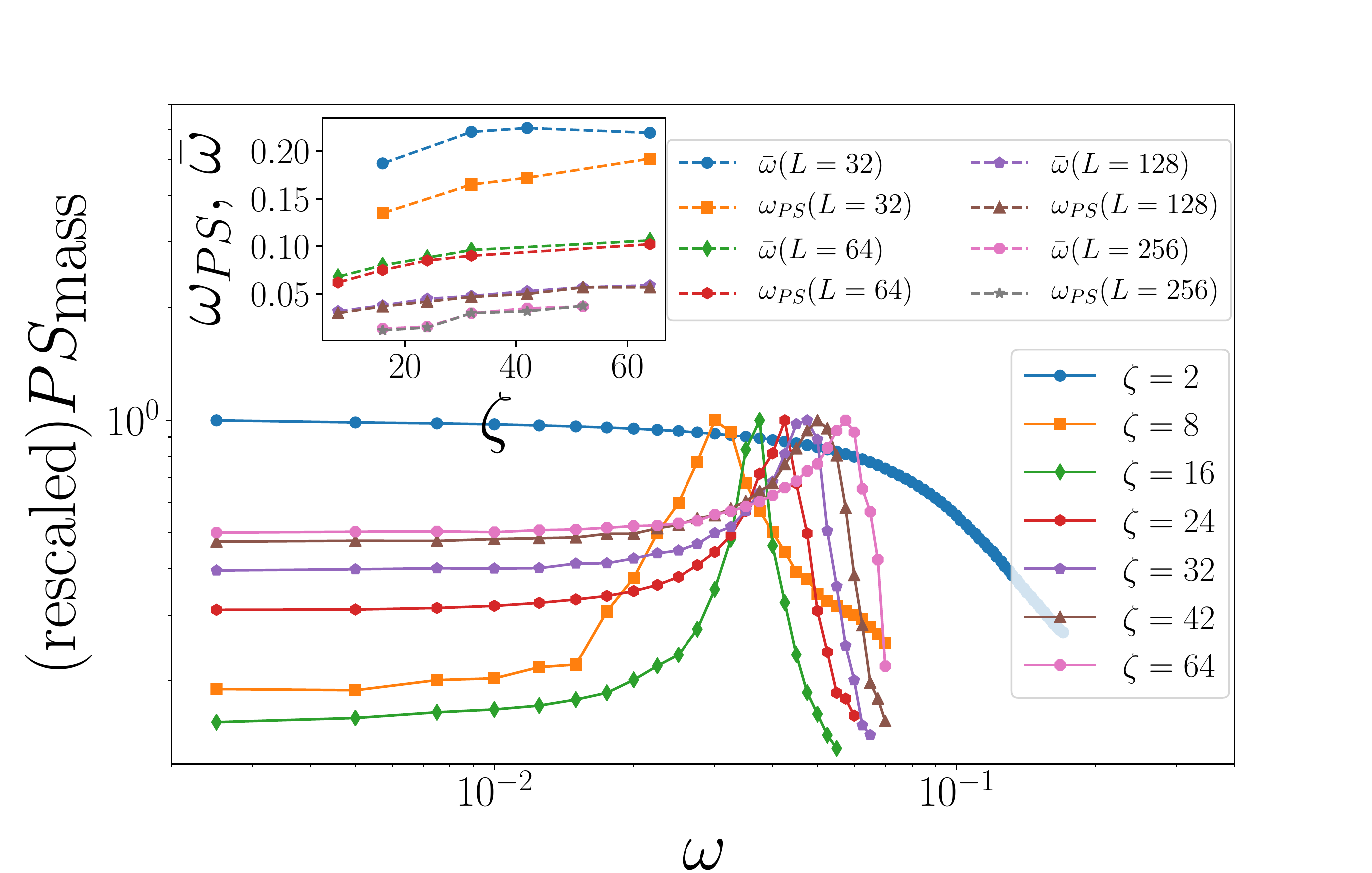}
		\caption{}
		\label{fig:PS_mass}
	\end{subfigure}
	\caption{(Color online): (a): The (shifted) energy average in terms of $t$ (the number of injection) for various rates of $\zeta$. Inset: $\bar{E}_1-\bar{E}_2$ in terms of $\zeta$. (b) Mass, as a function of time $T$ for various rates of $\zeta$ and $L$. (c) The time dependence of mass autocorrelation function. (d) The mass power spectrum in terms of frequency, and their peaks (inset). All figures are for $L=256$.}
	\label{fig:mass0}
\end{figure*}

The resulting samples have been shown in Figs~\ref{fig:samples1} and~\ref{fig:samples2}. It is seen that the configurations become smoother under the application of fillips. To test our predictions, we have calculated and plotted $\bar{E}$ in terms of $T$ for various rates of $\zeta$ in Fig.~\ref{fig:a_ave}. Two separate phases are seen in this figure: in the primitive times it varies linearly, and for large enough times it enters a new phase, e.g. for $\zeta=0$ it is nearly constant. However, for non-zero $\zeta$s we see that the prediction of MF is confirmed and some oscillations arise in which the average density increases with $T$. In the inset of this figure we have plotted the difference between these two value (among which $\bar{E}$ oscillates) $\bar{E}_1-\bar{E}_2$ in terms of $\zeta$, which quantifies these oscillations. Additionally, this energy drop becomes independent of $\zeta$ for large enough $\zeta$s, as predicted by MF calculations. This function starts from zero in small enough $\zeta$s and grows rapidly, and saturates in some $L$-dependent $\zeta^*$ which is served as the bifurcation point. In the following, we see that the physics of these two regimes are different.\\
In the MF approach, we claimed that these oscillations are due to two branches in Eq.~\ref{Eq:meanfield}. The Fig.~\ref{fig:mass_T} visualizes this event, in which a new characteristic reference point appears for large enough $\zeta$s. In this figure we have shown the mass of the avalanches ($\equiv$ the number of distinct toppled sites in the avalanche) as a function of time for various rates of $\zeta$ and $L$. Consider for example $L=64$ in the regime $\zeta\gtrsim 8$, for which the mass of some avalanches reach the system size, i.e. the top points in the figure whose mass is $(64)^2=4096$. These avalanches are the mentioned SAs, that are absent in small $\zeta$s. The SAs and the abrupt drop of average energy occur simultaneously, and therefore have the same origin (both belong to the lower branch of Eq. \ref{Eq:meanfield}). The avalanches that belong to the first branch, whose mean size grow linearly with the injections are called \textit{deformed avalanches (DA)}. The mean size of DAs depends on $E_{th}-\bar{E}$. \\
Although these figures are helpful for understanding the phenomenon, some other tests are necessary to quantify it. The auto-correlation function of the mass noise $\left\lbrace m(T)\right\rbrace_{T=1}^{T_{\text{max}}}$ (in which $m(T)$ is the mass of $T$th avalanche, and $T_{\text{max}}$ is the maximum $T$ in our analysis) is defined as:
\begin{equation}
f_{\text{mass}}(T_0)\equiv \left\langle  m(T)m(T+T_0)\right\rangle_{T} -\left( \left\langle  m(T)\right\rangle_{T} \right)^2
\end{equation}
in which the $T$-average of an arbitrary statistical observable is defined by~\cite{Davidsen} $\left\langle O\right\rangle_{T} \equiv\frac{1}{T_{\text{max}}}\sum_{T=0}^{T_{\text{max}}} O(m(T))$. Figure~\ref{fig:A_mass} shows the re-scaled autocorrelation function of $m(T)$ defined by $A_{\text{mass}}(T)\equiv \left( \left\langle  m(T)\right\rangle_{T} \right)^{-2}f_{\text{mass}}(T)$. Also the power spectrum is defined by
\begin{equation}
PS_{\text{mass}}(\omega)\equiv \lim_{T_{\text{max}}\rightarrow\infty}\frac{1}{T_{\text{max}}}\left| \int_0^{T_{\text{max}}} dT m(T)\exp\left[-i\omega T\right] \right|^2 
\end{equation}
and is proportional to the Fourier transform of $A_{\text{mass}}(T)$. The same definitions hold also for the activity noise.\\

As is seen in the Fig.~\ref{fig:A_mass}, as $\zeta$ increases (more precisely for $\zeta\gtrsim 8$) , some oscillatory behaviors appear. These oscillatory behaviors are due to the new reference point that was mentioned above. In the other words the oscillations in $\bar{E}(T)$ is responsible for this oscillatory autocorrelation. The most robust (long range) oscillations are found for $\zeta\approx 16$ for $L=256$. It is notable that these $\zeta$ values are $L$-dependent and are therefore non-universal. To strengthen the connection of these oscillations and the oscillations for $\bar{E}(T)$, one should calculate the power spectrum of the noise (here the mass) and find its peak. The resulting frequencies can then be compared with the frequencies of the oscillations of $\bar{E}(T)$. This is done in Fig.~\ref{fig:PS_mass} for $L=256$. For small $\zeta$ values ($\zeta=2$ in this figure), one finds no peak, whereas for larger $\zeta$s a peak appears which run with $\zeta$. In the inset, the position of these peaks ($\omega_{PS}$) has been shown for some lattice sizes, along with the average angular frequency ($\bar{\omega}$) obtained from the oscillations of $\bar{E}(T)$. The fact that these frequencies are properly matched show that they have the same origin, i.e. the system oscillates between two states: DAs and SAs. These frequencies increase monotonically with $\zeta$, revealing that the fillips facilitate the creation of the SAs.\\

\begin{figure*}
	\centering
	\begin{subfigure}{0.49\textwidth}\includegraphics[width=\textwidth]{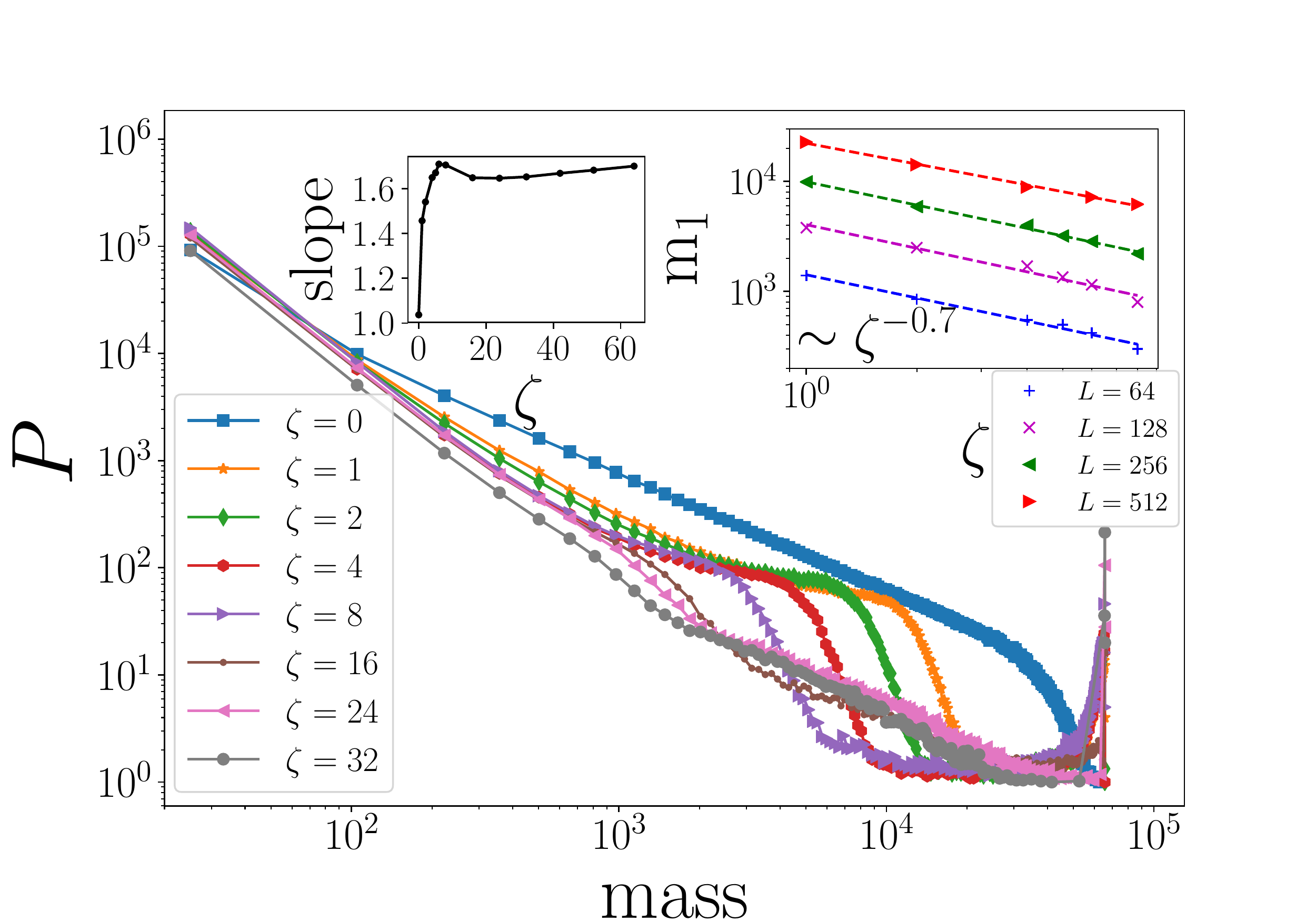}
		\caption{}
		\label{fig:P_mass}
	\end{subfigure}
	\begin{subfigure}{0.49\textwidth}\includegraphics[width=\textwidth]{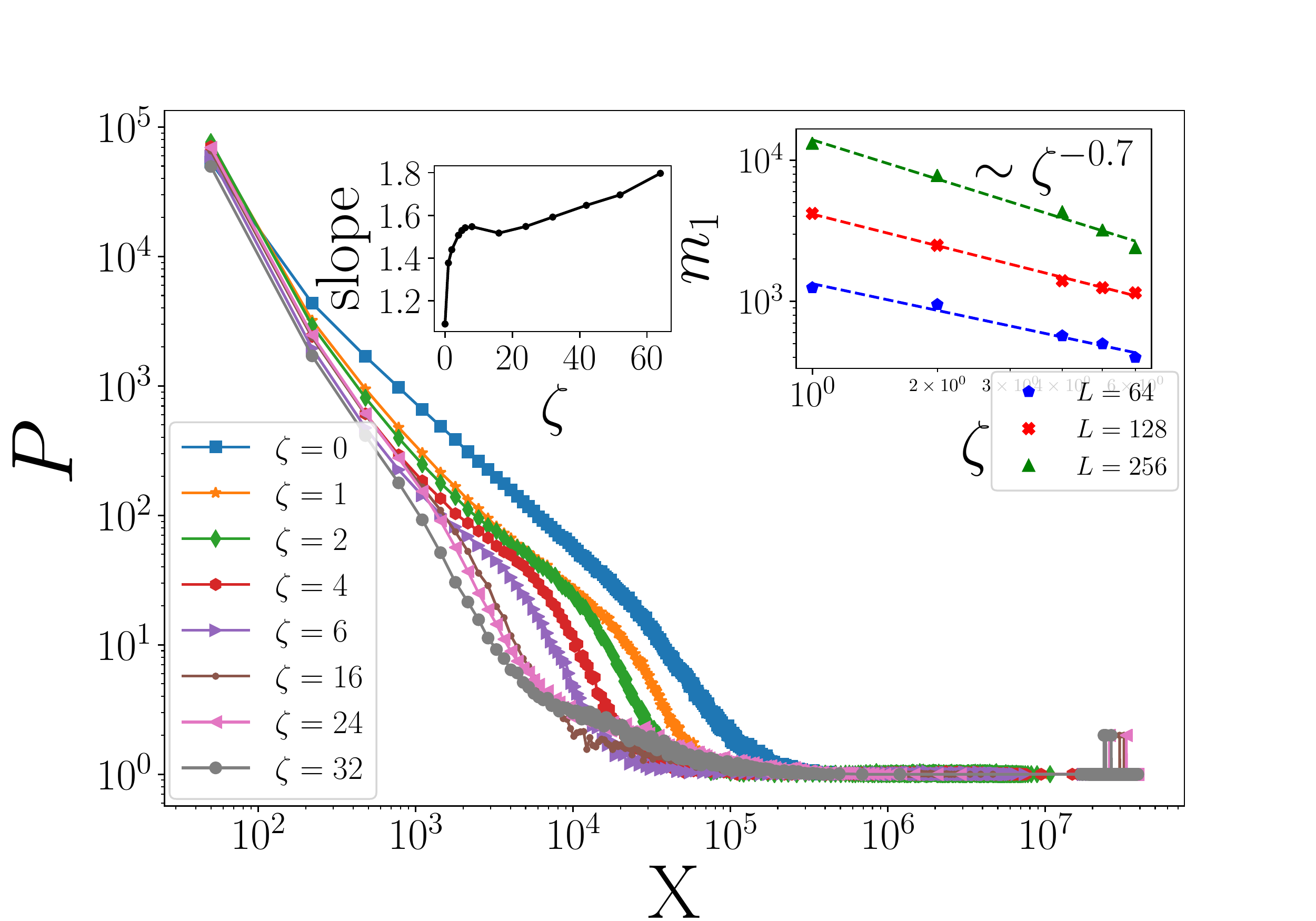}
		\caption{}
		\label{fig:P_X}
	\end{subfigure}
	\begin{subfigure}{0.49\textwidth}\includegraphics[width=\textwidth]{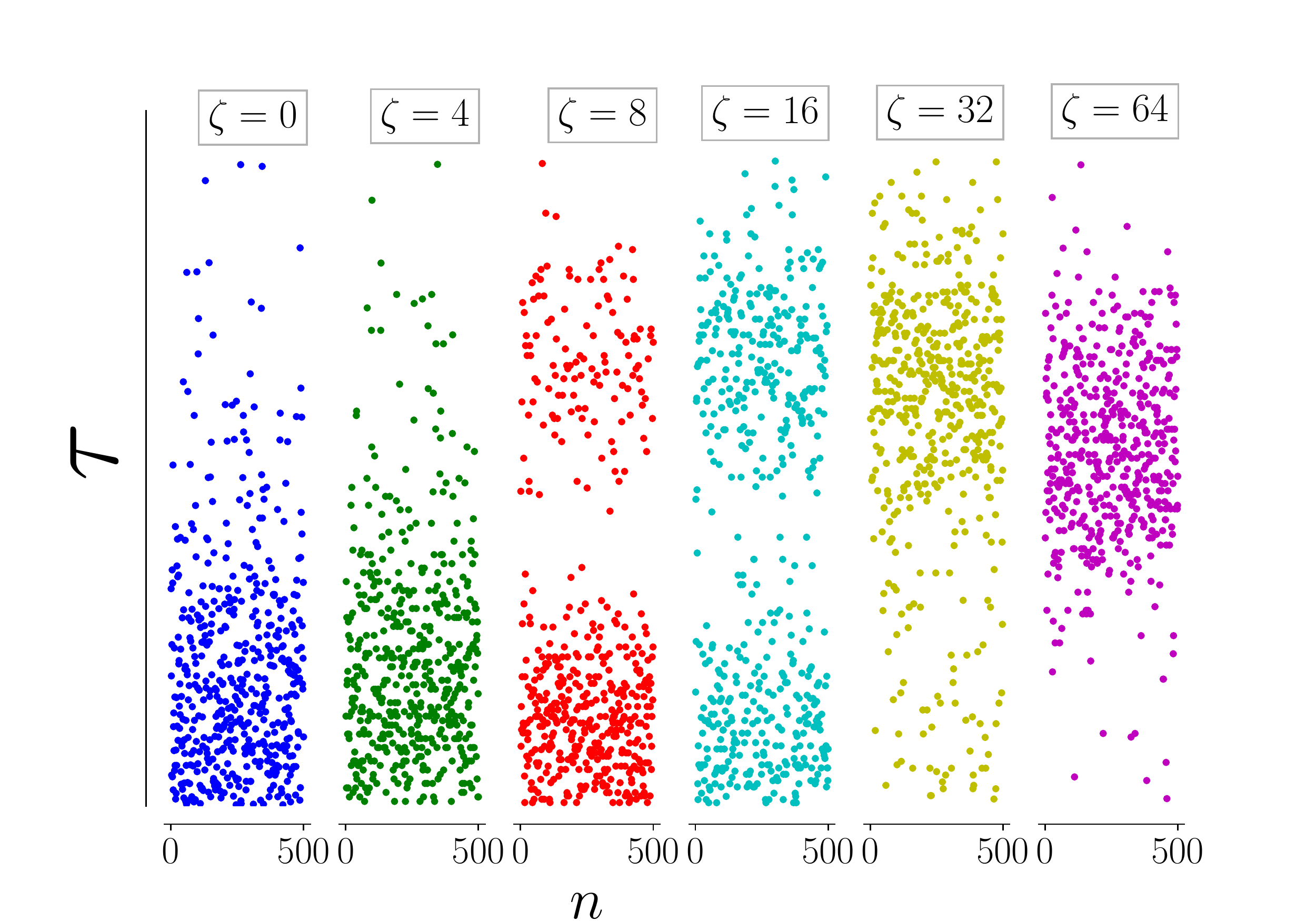}
		\caption{}
		\label{fig:tau_n}
	\end{subfigure}
	\begin{subfigure}{0.49\textwidth}\includegraphics[width=\textwidth]{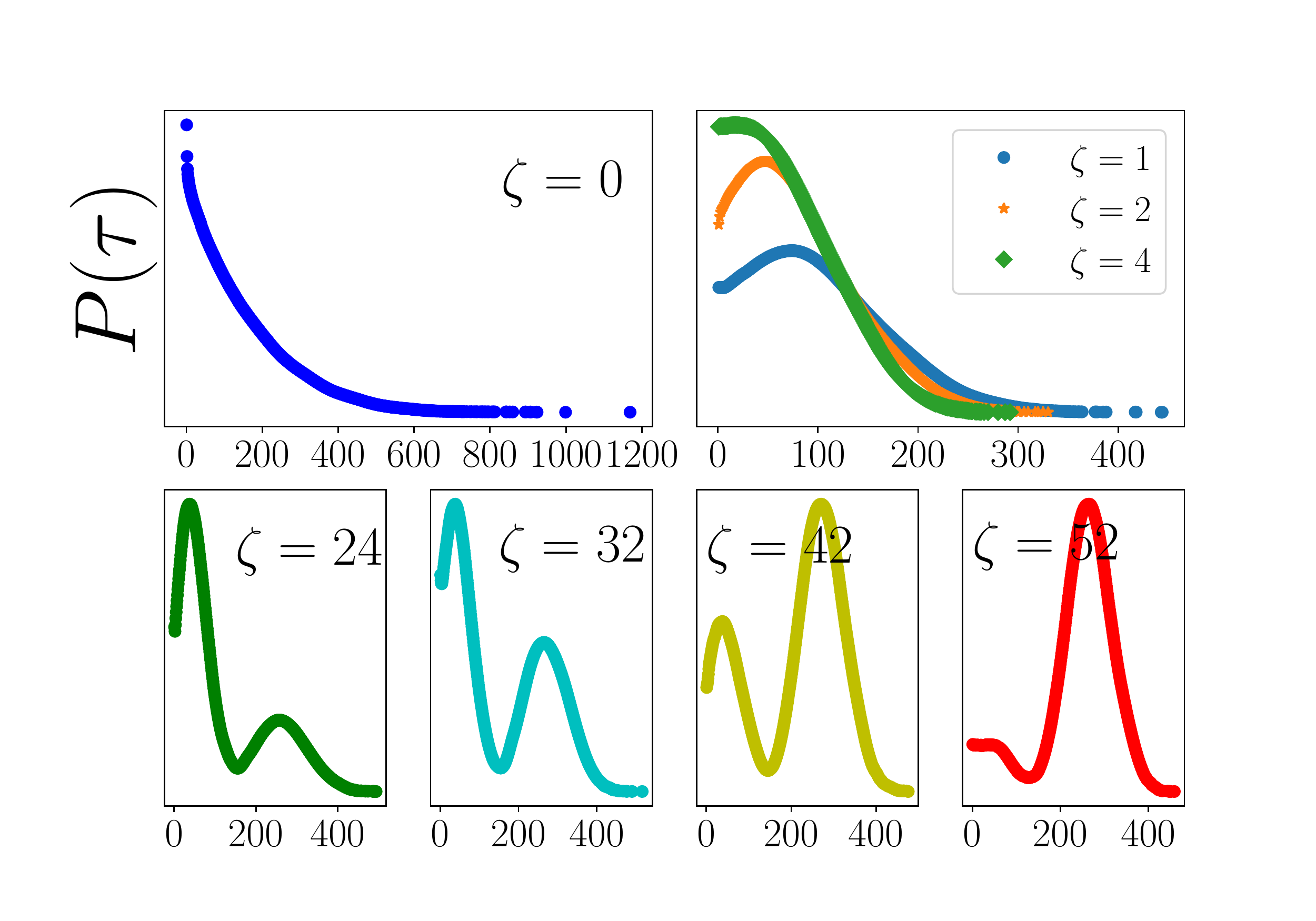}
		\caption{}
		\label{fig:P_tau}
	\end{subfigure}
	\caption{(Color online) The distribution of the mass (a) and the activity (b). The time series of REWT (c), and its corresponding distribution (d).}
	\label{fig:distribution}
\end{figure*}
The important quantities that reflect the state of the system in hand are the distribution functions (Figs.~\ref{fig:P_mass} and \ref{fig:P_X}. For $\zeta=0$, a power-law behavior is seen in Figs.~\ref{fig:P_mass} as expected, whereas for non-zero $\zeta$s the trend of the graphs change considerably (for $L=256$). Especially a new mass scale (namely $m_1$) comes to play bellow which the linearity of the log-log graph is retained. The slope of the graphs in small masses has been shown in the left inset of this graph. Also a sharp peak is seen at $\text{mass}=65536=(256)^2$, which are SAs as stated above. A more precise look at this figure reveals that $m_1$ decreases (in a power-law fashion) with increasing $\zeta$ up to $\zeta_c\approx 16$ above which it saturates. The same features are seen for the distribution of the activity $P(X)$ (Fig.~\ref{fig:P_X}).\\
The rare events waiting time (REWT) is another important stochastic quantity that has deep connections with the activity noise. It is defined as the waiting time between large avalanches, i.e. the avalanches with sizes larger than a threshold value $s_{\text{threshold}}$ (as a large event) which is fixed to be $2L^2$. We define REWT, denoted by $\tau(n)$, as the time interval between two successive rare events $s(T_{n})$ and $s(T_{n+1})$. In other words if the $(n)$th rare event occurs at time $T_{n}$ (i.e. $s(T_{n})>s^{\text{threshold}}$) and the next rare event occurs at time $T_{n+1}$, then $\tau(n)\equiv T_{n+1}-T_{n}$. We have observed that the results are nearly independent of $s_{\text{threshold}}$. Figures~\ref{fig:tau_n} and~\ref{fig:P_tau} and show the results for REWT for $L=256$. It is interestingly seen from the Fig.~\ref{fig:tau_n},  that for $8 \lesssim\zeta\lesssim 16$ the aggregation of events is on two reference points, i.e. the corresponding distribution functions are doubly peaked, whereas for $\zeta\lesssim 8$ they are singly peaked. This can better seen in Fig.~\ref{fig:P_tau}, in which the lower graphs correspond to $\zeta>8$. We see that, as $\zeta$ increases, the second peak grows, whereas the first one weakens. This is the signature of the first-order transition, although some other evidences are necessary.\\
In Summary, by \textit{smoothening} of (applying fillips on) the BTW system, the system meets various phases and various new characteristic scales. For $0\leq \zeta \lesssim \zeta^*$, although the range of avalanches decrease with $\zeta$ (see Figs.~\ref{fig:a_ave} and \ref{fig:mass_T} in which the distribution of large avalanches are lower than small avalanches), but no oscillations are found. In this interval a new mass scale ($m_1$ in Figs.~\ref{fig:a_ave} and \ref{fig:mass_T} which decreases in a power-law fashion with $\zeta$) is found which corresponds to the new $\tau$ scale in Fig.~\ref{fig:P_tau}. A bifurcation takes place in $\zeta=\zeta^*$, in such a way that for $\zeta^*\lesssim \zeta \lesssim\zeta_c$ some long-range oscillations are found (Fig.~\ref{fig:A_mass}). As $\zeta$ increases further, for $\zeta_c\lesssim \zeta$, $\bar{E}_1-\bar{E}_2$ (Fig.~\ref{fig:a_ave}) as well as the mass scale $m_1$ (Fig.~\ref{fig:P_mass} and \ref{fig:P_X}) saturate. At this interval a new $\tau$ scale appears and grows with $\zeta$. It is notable that these critical $\zeta$s are $L$-dependent and for $L=256$, $\zeta^*\approx 8$ and $\zeta_c\approx 16$.

\section*{Discussion and Conclusion}
\label{sec:conc}
According to these findings the try of smoothening of a sandpile cause the system to meet some unexpected phases. This smoothening translates to some other relaxation procedures in the other self-organized critical systems. For example, one may test the effect of deliberate artificial bursts (artificial earthquakes) on the time series of natural earthquakes. Our calculations demonstrate that the state of the underlying system depends on the strength of this smoothening, and the critical strengths depend on the system size. This smoothening cause some energies to accumulate in the system over a time interval, leading to some spanning avalanches, and also some oscillatory behaviors.

\end{document}